\documentclass{article}

\usepackage{PRIMEarxiv}

\usepackage[utf8]{inputenc} 
\usepackage[T1]{fontenc}    
\usepackage{hyperref}       
\usepackage{url}            
\usepackage{booktabs}       
\usepackage{amsfonts}       
\usepackage{nicefrac}       
\usepackage{microtype}      
\usepackage{lipsum}
\usepackage{fancyhdr}       
\usepackage{graphicx}       
\usepackage{xcolor}
\usepackage{caption}
\usepackage{placeins}
\usepackage{multicol} 
\usepackage{amsmath}
\usepackage{enumitem}
\usepackage{tabularx} 

\newcommand{\eqnref}[1]{Eq.~\ref{#1}}
\newcommand{\figref}[1]{Figure~\ref{#1}}
\newcommand{\tabref}[1]{Table~\ref{#1}}
\newcommand{\secref}[1]{Section~\ref{#1}}

\graphicspath{{media/}}     

\title{Optimizing Sales Forecasts through Automated Integration of Market Indicators
}

\author{
  Lina Döring \\
  Bielefeld University of Applied Sciences and Arts  \\
  Gütersloh, Germany  \\
  \texttt{lina.doering@hsbi.de}
   \And
  Felix Grumbach  \\
  Bielefeld University of Applied Sciences and Arts \\  
  Gütersloh, Germany 
  \And
  Pascal Reusch  \\
  Bielefeld University of Applied Sciences and Arts \\  
  Gütersloh, Germany 
}

\begin{document}
\maketitle

\begin{abstract}
Recognizing that traditional forecasting models often rely solely on historical demand, this work investigates the potential of data-driven techniques to automatically select and integrate market indicators for improving customer demand predictions. By adopting an exploratory methodology, we integrate macroeconomic time series, such as national GDP growth, from the \textit{Eurostat} database into \textit{Neural Prophet} and \textit{SARIMAX} forecasting models. Suitable time series are automatically identified through different state-of-the-art feature selection methods and applied to sales data from our industrial partner. It could be shown that forecasts can be significantly enhanced by incorporating external information. Notably, the potential of feature selection methods stands out, especially due to their capability for automation without expert knowledge and manual selection effort. In particular, the Forward Feature Selection technique consistently yielded superior forecasting accuracy for both SARIMAX and Neural Prophet across different company sales datasets.
In the comparative analysis of the errors of the selected forecasting models, namely Neural Prophet and SARIMAX, it is observed that neither model demonstrates a significant superiority over the other.
\end{abstract}


\begin{multicols}{2}
\section{Introduction}
This section provides a practically oriented motivation, a brief overview of related research, and outlines the primary contributions that this paper aims to make.

\subsection{Motivation and Industrial Use Case}
Time series modelling and forecasting plays a crucial role in a range of fields, from climate modelling to electricity planning and business decision-making \cite{bennettAutoregressiveExogenousVariables2014, yerlikayaClimateChangeForecasting2020, ensafiTimeseriesForecastingSeasonal2022, wanMultivariateTemporalConvolutional2019, xiaoAFSTGCNPredictionMultivariate2022}. Businesses heavily rely on forecasting to plan operations, manage resources, and respond to market dynamics \cite{triebeNeuralProphetExplainableForecasting2021}. Foremost, forecasting is inherently uncertain. No model can predict the future with absolute certainty because it depends on various often unknown factors \cite{castleRobustApproachesForecasting2015}.
Additionally, conventional forecasting methods often focus solely on historical data of the time series it wants to predict, overlooking external influences. Businesses normally are affected by various external factors, such as market effects or government decisions. Incorporating additional time series, containing relevant information regarding the market, can enhance forecasting performance \cite{tyralisLargescaleAssessmentProphet2018}. However, identifying and quantifying these dependencies is usually very challenging and associated with a lot of effort \cite{bennettAutoregressiveExogenousVariables2014, borkinAddingAdditionalFeatures2019, chenMultiScaleAdaptiveGraph2023, wuConnectingDotsMultivariate2020}.
Rather than striving for perfect certainty in identifying all dependencies, a more practical approach involves testing the influence of several external time series on the forecasting quality \cite{hallCorrelationbasedFeatureSelection1999}. Businesses can continuously assess what factors enhance their predictions. Thereby, insights of the influence of specific external time series on their demand patterns can be gained.
This work will look at the real case of a medium-sized German company producing cleaning supplies and associated machines for businesses. The forecast of future demand of their products should be improved by adding external information to the forecast. For instance, the demand for industrial machine cleaning supply may not only depend on past demand but also on the overall market conditions, e.g., the machines sales themselves.

\subsection{Related Work}
For the task of incorporating market indicators into forecasts, three questions are discussed in the related work: Which forecasting model to choose, what markets indicators also called exogenous variables to incorporate into the forecast and how to evaluate the forecast. In the world of business, making decisions and planning for the future relies on predicting what customers will need and effectively managing resources like materials and workforce accordingly. Forecasts are an important tool for planning under uncertainty \cite{jimenezMultiobjectiveEvolutionaryFeature2017}.
There exists a great variety of forecasting models. The landscape of time series forecasting models encompasses classical statistical models such as ARMA, ARIMA, and Exponential Smoothing \cite{ensafiTimeseriesForecastingSeasonal2022, yangNetworkTrafficForecasting2021, jimenezMultiobjectiveEvolutionaryFeature2017, hyndman2018forecasting, boxTimeSeriesAnalysis1976} as well as more contemporary machine learning models. The latter includes various neural network architectures, notably Long-Short-Term-Memory (LSTM), Recurrent Neural Networks (RNN), Convolutional Neural Networks (CNN)\cite{yangNetworkTrafficForecasting2021, ensafiTimeseriesForecastingSeasonal2022, xiaoDualStageAttention2021, parmezanEvaluationStatisticalMachine2019, limTimeseriesForecastingDeep2021} and a hybrid combination of statistical and machine learning methods \cite{bennettAutoregressiveExogenousVariables2014}.
Because of its real world's relevance, business related time series forecasting is well represented in the literature \cite{ensafiTimeseriesForecastingSeasonal2022, jimenezMultiobjectiveEvolutionaryFeature2017, wilmsInterpretableVectorAutoRegressions2017, bojerKaggleForecastingCompetitions2021}.
Forecasts typically rely on historical data, but real-world scenarios involve market indicators known as exogenous variables or regressors (These words will be used synonymously in this work). There are several approaches on how to incorporate exogenous time series into a forecasting model \cite{triebeNeuralProphetExplainableForecasting2021, xiaoDualStageAttention2021, jimenezMultiobjectiveEvolutionaryFeature2017, wilmsInterpretableVectorAutoRegressions2017, castleRobustApproachesForecasting2015}.
As Wolpert et al. stated already 1997 in his “no free lunch theorem” - “there is not a model which will always perform better than other models" \cite{wolpertNoFreeLunch1997}.
Choosing the right model is not the key factor to improve forecast performance. However, taking into account exogenous variables will improve forecasting results \cite{tyralisLargescaleAssessmentProphet2018, jimenezMultiobjectiveEvolutionaryFeature2017, hongProbabilisticElectricLoad2016}.
That is why this work only focuses on the selection of market indicators. Two models will be used to have a broader comparison, especially on how exogenous variables are incorporated by the methods. The literature suggest comparing a naive one and a more advanced approach in such cases \cite{tyralisLargescaleAssessmentProphet2018, hyndman2018forecasting} 
Therefore a statistical method and a more advanced but still interpretable holistic method are selected for the forecasting task. Both models can incorporate exogenous variables:

\begin{itemize}[leftmargin=*]
	\item \textbf{SARIMAX} (AutoRegressive Integrated Moving Average with eXogenous variables) from Python 
\verb|statsmodels| library \cite{SARIMAXIntroduction2023}
	\item \textbf{NeuralProphet (NP)} model from a library from the research group surrounding Oskar Triebe at Standfort University \cite{triebeNeuralProphet2021}
\end{itemize}

The question already motivated above is the identification of meaningful exogenous information or time series to incorporate into a forecast. For demand forecasting, for example, macroeconomic variables are used to set the forecasting within the broader context of its associated business environment  \cite{castleRobustApproachesForecasting2015}.
But how to find market indicators that are actually relevant for the forecast and therefore improve its results is not a topic of many articles. Tyralis et al. recommend to focus on this selection of the exogenous variables in future research \cite{tyralisLargescaleAssessmentProphet2018}. Some approaches include external factors according to domain knowledge and therefore a manual selection \cite{triebeNeuralProphetExplainableForecasting2021}. Further suggestions to select those variables are partial mutual information or genetic programming \cite{tranSelectionSignificantInput2015}. The identification of exogenous variables that improve the forecast is a feature selection task \cite{jimenezMultiobjectiveEvolutionaryFeature2017}.
The most common way to compare and evaluate forecasts is to hold back a sequence to evaluate the performance of the model out-of-sample. Most commonly this sequence is in the end of the time series and is as long as the forecasting horizon. 
The difference between the forecasts of the model and the test data for this time range is used to calculate an error measure to compare different forecasts for this time range with each other \cite{bojerKaggleForecastingCompetitions2021}.
 
\subsection{Contribution} \label{chapter:contribution}
The focus of this work is optimizing time series forecasting for the demand of our industrial research partner using market indicators as exogenous variables. Selecting the right indicators to improve the forecasting in a structural manner, instead of choosing the variables solely based on domain knowledge, is a new approach. The goal is to enhance forecasting performance and provide practical insights for real-world applications. The research questions (RQ) are:

\begin{itemize}[leftmargin=*]
	\item \textbf{RQ1}: How can real-world sales forecasts be optimized by integrating exogenous variables, explicitly market indicator time series from Eurostat (Statistical office of the European Union)?
	\item \textbf{RQ2}: How can suitable market indicators be identified systematically using feature selection techniques?
	\item \textbf{RQ3}: How do predictions of the two forecasting models SARIMAX and NP differ when integrating exogenous variables?
\end{itemize}
	
\section{Methods and experimental set up} 
This section begins by explaining the main facts and characteristics of the chosen time series forecasting algorithms. It then discusses the data to be used and how it will be pre-processed.
Next, all variable selection methods, that are performed for this work, are presented. Finally, the experimental setup is outlined.

\subsection{Time series Forecasting Algorithms} \label{sec:ts-forecasting-algos}
The domain of effective planning necessitates a profound understanding of the future. But given the intrinsic uncertainty of forthcoming events, this future needs to be predicted. Planning is inherently reliant on historical data, ideally in the structured format of time series, which chronicles temporal sequences of observations.
For this work, the forecasts needed are going to be modelled by the time series forecasting approaches SARIMAX and NP. One can consider SARIMAX or its predecessors like ARMA and ARIMA \cite{boxTimeSeriesAnalysis1976} as old and therefore not relevant any more but literature proofs, they are still considered relevant \cite{nokeriForecastingUsingARIMA2021, moroffMachineLearningStatistics2021}. This work wants to incorporate exogenous variables in forecasting models. SARIMAX and NeuralProphet are able to perform this task and are commonly used for it \cite{tyralisLargescaleAssessmentProphet2018, bennettAutoregressiveExogenousVariables2014, parmezanEvaluationStatisticalMachine2019, triebeNeuralProphetExplainableForecasting2021}. 
Additionally, this work does not try to find the best model to forecast the use case companies demand, but to find a way on how to support this forecast with market indicators. Furthermore, this choice is motivated by practicability of the results: it should remain interpretable to be useful for practitioners. 

\subsubsection{SARIMAX} \label{chapter:sarimax}
The SARIMAX model is an extension of the ARIMA model, which was first introduced by Box and Jenkins in 1976 \cite{boxTimeSeriesAnalysis1976}.
It is an acronym for \textbf{S}easonal \textbf{A}utoregressive \textbf{I}ntegrated \textbf{M}oving \textbf{A}verage with e\textbf{X}ogenous Variables, which is a powerful time series forecasting model that incorporates various components to capture and predict complex temporal patterns in data.

\textbf{AR (Auto-regression)}: ARIMA, the basis of SARIMAX, relies on auto-regression, denoted as 'AR.' This component involves making forecasts by linearly combining past values, often referred to as 'p-lags' as it can be seen in \eqnref{eq:AR}. In other words, ARIMA examines the historical values of the time series and uses a weighted combination of these past observations to predict values $\hat{y}_t$. The weights that are assigned to the lags are the corresponding $\alpha$-values. 
\begin{equation}
	\hat{y}_{t} = \alpha_{t-p}y_{t-p} + \alpha_{t-p+1}y_{t-p+1}+...+\alpha_{t-1}y_{t-1} 
	\label{eq:AR}
\end{equation}

\textbf{MA (Moving Average)}: The 'MA' in ARIMA stands for Moving Average. This aspect signifies that ARIMA's forecasts depend on the errors made in past forecasting values. However, these forecasting errors can only be determined after the model is fitted, making it challenging to estimate the optimal number of past errors to consider, which are also known as 'q-lags' (see \eqnref{eq:MA}). In parallel with $\alpha$ for the AR-component, the $\theta$-values are the weights of the lags $\epsilon$.
\begin{equation}
	\hat{y}_{t} = c + \theta_{t-q}\epsilon_{t-q} + \theta_{t-q+1}\epsilon_{t-q+1} + ... + \theta_{t-1}\epsilon_{t-1}
	\label{eq:MA}
\end{equation}

The error $\epsilon$ is calculated as seen in \eqnref{eq:error-ma}
\begin{equation}
	\epsilon_t = y_t - \hat{y}_t 
	\label{eq:error-ma}
\end{equation}

So far a so called ARMA model is defined but when it should be used to model time series, it needs to be stationary. Because the ARMA model has a strong stationary assumption. A stationary time series is a type of time series data where statistical properties, such as the mean, variance, and autocorrelation, remain constant over time. In simpler terms, it is a time series in which the underlying data generating process does not show significant changes or trends. Stationarity is an essential concept in time series analysis and forecasting because many time series models assume or work best with stationary data. Therefore, the data needs to be transformed into stationary data before modelling it with the ARMA model.
A common approach to make data stationary is by integrating.

\textbf{I (Integration)}: The 'I' in SARIMAX stands for Integration. In contrast to ARMA models, ARIMA models include this component to address the stationary assumption. To make non-stationary data stationary, SARIMA employs differencing or integrating within the model. Differencing is often represented as a triangle or delta symbol ($\Delta$) in the formula (see \eqnref{eq:arima-full}).
\begin{align}
	\Delta Y_{t} = c + \biggl( \sum_{i=1}^{p} \alpha_{i} \Delta y_{t-i} \biggr)
    - \biggl( \sum_{i=1}^{q} \theta_{i} \epsilon_{t-i} \biggr)+ \epsilon_t 
    \label{eq:arima-full}
\end{align}

\textbf{S (Seasonal)}: SARIMA also incorporates seasonal effects, denoted by 'S.' This component acknowledges the presence of seasonal patterns in time series data, which are common in various fields, especially when dealing with human behavior. Examples of seasonality include the demand for winter coats increasing in colder months but declining in the summer. Seasonal patterns are commonly weekly or monthly, but can also manifest in other intervals \cite{boxTimeSeriesAnalysis1976, parmezanEvaluationStatisticalMachine2019}. See in \eqnref{eq:seasonal-component} how the seasonal component is build.
\begin{equation}
\left( \sum_{j=1}^{P} \phi_{j} \Delta_s y_{t-j s} \right) - \left( \sum_{j=1}^{Q} \Theta_{j} \epsilon_{t-j s} \right) + \epsilon_t
\label{eq:seasonal-component}
\end{equation}

\textbf{X (Exogenous Variables)}: The 'X' in SARIMAX shows the model's ability to consider exogenous variables when making forecasts. Exogenous variables are external factors that can influence the time series. To accommodate these variables, SARIMAX includes an additional term $\beta X$. This term is added to the basic ARIMA model (see \eqnref{eq:arima-full}), enabling the incorporation of exogenous information into the forecasting process. So the $X$ is integrated by estimating its influence or coefficients to the model while fitting. This is typically done with maximum likelihood estimation or least squares estimation. These coefficients are considered as constants and therefore not depending on $t$ \cite{vagropoulosComparisonSARIMAXSARIMA2016}. The full model, including the seasonal component and $\beta X$ from \eqnref{eq:seasonal-component} can be found in \eqnref{eq:full-model}.
\begin{align}
\Delta Y_{t} &= c + \beta X_t + \left( \sum_{i=1}^{p} \alpha_{i} \Delta y_{t-i} \right) - \left( \sum_{i=1}^{q} \theta_{i} \epsilon_{t-i} \right) \nonumber\\
&\quad + \left( \sum_{j=1}^{P} \phi_{j} \Delta_s y_{t-j s} \right) - \left( \sum_{j=1}^{Q} \Theta_{j} \epsilon_{t-j s} \right) + \epsilon_t
\label{eq:full-model}
\end{align}

For also including exogenous variables in an out-of-sample forecast, the exogenous variable future values need to be known. As it normally is not known, a workaround of making an easy forecast (by linear regression) for each regressor and adding this to the prediction was set up.
All the parameter above are written in the model as seen in \eqnref{eq:sarimax-letters}. $p$ stands for the number of past lags that should be taken into account, $d$ for the degree of differentiation, $q$ for the number of past errors that should be incorporated into the time series modelling. This three parameter are also called the order of the ARIMA model. The next parameters are the seasonal order. $S$ stands for the number of seasons faced during a year, e.g. 12 means monthly periodicity. For $P$, $D$, $Q$ the time series is shifted by $s$. The best number of past lags, errors and degree of differentiation when adding this time shift of $s$ to the time series can be found \cite{parmezanEvaluationStatisticalMachine2019}.
\begin{equation}
	SARIMAX(p,d,q)(P,D,Q)_s
	\label{eq:sarimax-letters}
\end{equation}

\begin{figure*}
	\centering
	\includegraphics[width=1\linewidth]{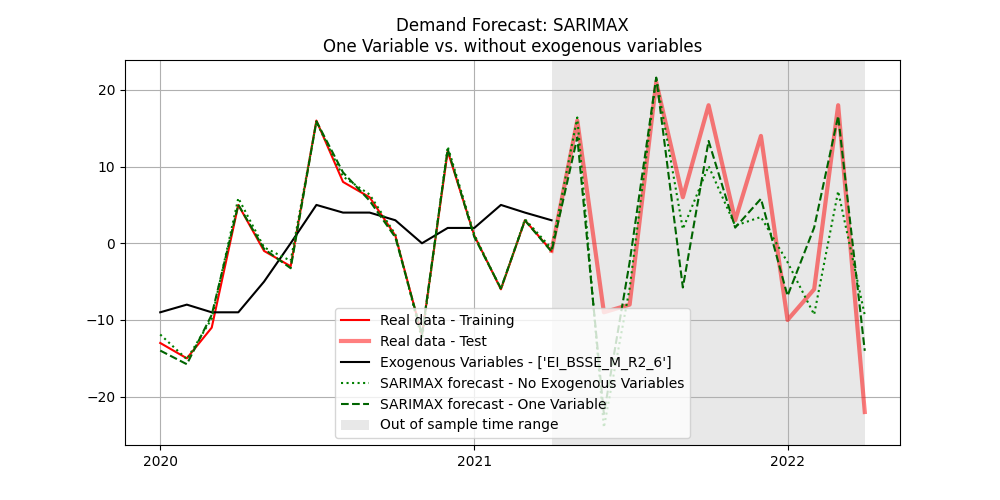}
	\captionsetup{font=small}
	\caption{This figure shows forecasts of \textbf{SARIMAX without and with one exogenous variables}. Both models were trained on the normalized BC1\_64 data. One with adding an exogenous variable from Eurostat dataset. This plot only shows the time range from 2020 on to have better visibility on the out-of-sample behavior of the forecasts (2021.05-2022.04). The values for the exogenous variable (in black) are only shown for the training data time range (until 2021.04) as these are the values taking into account from the model. All plots in this work are self-generated with the Python library \texttt{matplotlib} or MS PowerPoint.} 
	\label{fig:forecastsarimaxone}
\end{figure*}

\figref{fig:forecastsarimaxone} shows how the SARIMAX model behaves differently when adding an exogenous variable. The following effects can be seen: In comparison with the model without an exogenous variable, the in-sample performance does not change at all (2020.01-2021.04). For the out-of-sample forecast, different behavior is observed. For example, the upward trend of the exogenous variable (shown for the in-sample time range in black) is visible in a slight stronger positive trend of the forecast incorporating an exogenous time series, especially for the last peak at 2022.03.
The market indicator used as an exogenous variable here is a monthly business indicator for services. This indicator is build based on business situation, demand, perceived economic uncertainty, employment and selling prices of a representative group of companies in the service sector \cite{ServicesMonthlyData, MetadataBusinessConsumer}. This indicator was selected from one of the methods in this work to optimize the SARIMAX forecast in comparison to a forecast without exogenous variables. This will be explained in detail later in this work. It is not intuitively clear why such an indicator of the service sector is optimizing the forecast for industrial cleaning supply. It can be randomly e.g. that it is supporting the forecast because the trend of this indicator is quite similar to the one of the products sales.

\subsubsection{Neural Prophet} \label{chapter:np}
NP is a time series forecasting model introduced by Triebe et al. 2021. It aims to provide accurate, understandable forecasting without the need for extensive hyperparameter tuning \cite{triebeNeuralProphet2021}.
In business decision-making, the choice of relatively simple forecasting models holds importance. This allows decision makers to understand why a model produces a particular forecast. Understanding the forecasts is crucial in business decision-making \cite{triebeNeuralProphetExplainableForecasting2021}.
The authors of the former Prophet model \cite{taylorForecastingScale2018} and the more recent NP model followed the idea of separating time series generated by human behavior, like demand time series, into their different parts that are intuitively understandable. The resulting general calculation is shown in \eqnref{eq:np-general} \cite{taylorForecastingScale2018, triebeNeuralProphetExplainableForecasting2021}.

\begin{equation}
	\hat{y}_t = T_t + S_t + E_t +  F_t + A_t + L_t
	\label{eq:np-general}
\end{equation}

The $T_t$ function should model the trend of a time series. It is modelled with a piecewise linear trend function which changes the trend or gradient at change points that are either put automatically from the model or can be integrated from the user. $S_t$ represents the seasonality a time series might have. This is modelled with Fourier terms in the Prophet models.

The next parameter is the $E_t$. It models events that occur that have nothing to do with seasonalities (e.g. special fairs and promotions that have a demand impact for a company). This information can be incorporated into the model manually by introducing an event vector of equal length to the time series, with zero values except within the specific date range corresponding to the occurrence of the event.

The addition of $F_t$, which is for adding future known variables to the model, works similar to the event part.

Next up is the auto-regression of the time series, $A_t$ which models the influence of past values of the time series itself to the forecast. The order of lags can be set automatically by the model. The last parameter is the most interesting for this work.

It is $L_t$ which adds the exogenous variables to the model. This is also modelled autoregressively, and the model automatically takes the same number of lag as for the time series itself if not modelled differently \cite{taylorForecastingScale2018, triebeNeuralProphetExplainableForecasting2021}.

Because NP was build to be used by practitioners with no or not much knowledge about time series modelling, it also does pre-processing and hyperparameter tuning by itself automatically, if the according parameter is not set by the user \cite{triebeNeuralProphetExplainableForecasting2021, taylorForecastingScale2018}.
The model does pre-processing on its own by incorporating missing values with bidirectional linear interpolation, and it min-max normalizes the data by itself per default \cite{triebeNeuralProphetExplainableForecasting2021}. The reason for not using the former or more simple Prophet model is that it is not able to incorporate exogenous variables. There will be no use of the neural network options NP is offering for the auto-regression part of the model. Triebe, who is the main author of the NP paper, developed the Auto-regression net (AR-Net) for this. The simplest version of the AR-Net takes all lags as input values, one lag at each input neuron in the input layer, and has one output neuron per forecasting period in the forecasting horizon. This avoids the need of a new model for every further step down the forecasting horizon, as it is the case in normal auto-regression  \cite{triebeARNetSimpleAutoRegressive2019}. As mentioned previously, this was not used in this work's experiments, as this is no default setting of NP and not the focus of this work.
As this work aims to provide a solution that can be used by practitioners, the default settings of NP are kept for the experiments. 

\subsection{Datasets}
In this study, demand data is used from our industrial research partner, encompassing all product groups and market segments within its operations. The historic demand data is used to predict future demand.
To enhance the forecast, external time series data should be added. For this work, this external data is taken from Eurostat. 

\subsubsection{Demand Time Series} \label{internal-ts}
The demand data from our industrial research partner was collected monthly, since 2016, and was taken from the company's Enterprise Resource Planning (ERP) system.
The company is a cooperation partner in the research project "Smart Demand Forecasting". Aim of this project is to enhance sales forecasting.
At the start of the research project with the company, the data needed to be prepared. A data warehouse was build to load the demand data from the ERP system with an Extract-Transform-Load (ETL) process. The data was cleaned and harmonized to be available in a high quality. 
For example, an anomaly detection as well as a check for sampling frequency and a check for missing data were conducted.
It is essential to address anomalies cautiously, distinguishing between meaningful outliers (like the dip in many stock charts in 2020 caused by the Covid pandemic) and technical errors. Standardizing the sampling frequency, such as resampling to regular intervals, ensures data consistency \cite{CORTESIBANEZ2020385}.

The data warehouse is multidimensional  to be able to show the sales data in different dimensions for every month. Therefore, the sales data now can be separated and shown at different abstraction levels. For example, it can be drilled up and down on region, sector, customer group or single customer level. As well as on product or product group level. Hence, sales data can be analyzed on all the named dimensions at every sales date (monthly). This whole procedure also belongs to pre-processing of real world data. In \secref{sec:preproc-preselect} only the additional pre-processing of the data for the explicit experiments in this work is explained in detail.

Another example subproject within this research project involves employing meta-heuristics to pinpoint "leading" and "lagging" time series data, thereby offering a "Trend Detection System" for the company's specific use case. This "Trend Detection System" enables the identification and monthly reporting of products that perform exceptionally well or poorly within a designated time frame in a particular industry sector.

For this work, three product time series were chosen.
Decision criteria was that it has no zero values (month when zero units of the product were sold) and does not show a time series approximately remaining at the same level. The last criteria ensures that a more complex forecasting model than e.g. a simple linear regression is needed to predict the time series appropriately.
The three different product time series data sets are named BC1, BC2 and BC3 for this work.

\subsubsection{Market indicators}
In addition to the internal demand data, we incorporated external data and time series information sourced from the Eurostat data API. Eurostat offers a comprehensive database containing a wide array of statistical datasets pertaining to the European Union (EU). These datasets encompass diverse economic indicators, ranging from inflation rates and GDP growth to property prices but also indicators for consumer good trade or tourism in the member states of the EU. This data is accessible to the public \cite{Eurostat}, available via an API. It also exists a Python package named \texttt{eurostat} to read in the API data via python \cite{EurostatPythonPackage}.


\subsection{Data pre-selection and pre-processing} \label{sec:preproc-preselect}
As the demand data from our industrial research partner is captured monthly, all Eurostat datasets (7550 datasets) were filtered accordingly for monthly available data (down to 273 datasets). 
To refine the dataset's relevance to the business-focused analysis, a process of filtering for datasets with business-related implications was initiated. Each dataset within Eurostat is accompanied by specific parameters that provide contextual insights. These parameters contain information whether a dataset pertains for example a business trade indicator or external trade indicators. Subsequently, a manual review of all available parameters in the monthly datasets was conducted to determine the suitability of each dataset in enhancing the forecasting of industrial cleaning product sales. Afterwards, the datasets were filtered whether they contained one of the relevant parameters or not (133 datasets left). Given that the demand data spans from 2016 onwards, an additional criterion was applied to ensure that the selected exogenous time series possessed values dating back to 2016 as well as the demand data (48 datasets left). The data was gathered from Eurostat, end of September 2023. 

This is how 48 datasets including, 11944 time series were derived from Eurostat data. Viewing a few of these datasets closely already showed that the time series in one dataset often show similar courses but in different scales. Therefore, it is assumed that one time series per dataset will be enough to consider here. This also eased the task and saves computational resources. Additionally, this work is an explorative approach how to support a forecast with market indicators selected automatically and without the need for the forecaster to know all potential time series well. Therefore, we have 48 different time series as potential market indicators for the forecasting. Some of these time series are described later in the result section when they are selected as exogenous variables from some models, as it would be too extensive to have them all described in this work.

The demand data and the filtered external data from Eurostat are merged together. 
This merged data is pre-processed all-in-one by the data processing steps smoothing, min-max normalization, and discretization, with the removal of linear trends \cite{kotuDataScienceProcess2019, CORTESIBANEZ2020385}. 
It's worth mentioning that for the NP model, data normalization is not imperative, as the model inherently performs this task during its own pre-processing \cite{triebeNeuralProphetExplainableForecasting2021}.

\subsection{Variable parameter selection methods}
To answer RQ2, several variable selection methods were proposed. They are presented in this section.
As previously mentioned in the motivation section, the identification of market indicators that have an influence on the target time series to be forecasted, represents a challenge and one that most certainly will not yield definitive answers with absolute certainty.
To avoid this problem, the potential enhancement of forecasting performance resulting from the inclusion of various market indicators should be empirically assessed.
The literature provides diverse approaches for the variable selection task \cite{xiaoAFSTGCNPredictionMultivariate2022, wuConnectingDotsMultivariate2020, wilmsInterpretableVectorAutoRegressions2017, xiaoDualStageAttention2021, assafExplainableDeepNeural2019}.
When the problem is broken down, we have one variable to predict and other variables which may influence the result of the prediction. In such a situation, it needs to be decided which variables deliver relevant additional information and in what combinations, and which one does not. This can be identified as a feature selection task. The approaches that will be tested in this work are explained in this section.

\subsubsection{Correlation-Based Selection}\label{chapter:correlation-analysis}
A common and intuitive feature selection technique involves the selection of variables predicated on their correlation with the target variable, in this case, demand. Variables demonstrating a robust correlation with demand are deemed suitable for inclusion, given their anticipated influence on forecasting precision. Among the highly correlating variables, those displaying minimal mutual correlation are chosen to be added to the model \cite{hallCorrelationbasedFeatureSelection1999}.
The correlation-based analysis successfully identified exogenous variables, surpassing a designated 75\% threshold of minimum correlation among the 48 distinct candidates. To further refine this selection, variables correlating with each other the least were finally used as exogenous variables. It is imperative to avoid redundant variables, as their simultaneous use would not provide additional value. Consequently, another correlation test is conducted on the variables chosen from the 75\% threshold. Only those variables correlating to each other less than 30\% were ultimately incorporated into the model. Normally these thresholds are higher - 80\% and 20\%, but with these thresholds no correlating time series would have been found for most datasets \cite{gogtayPrinciplesCorrelationAnalysis2017}.

\subsubsection{LASSO Regression} \label{chapter:lasso-regression-explain}
Least absolute shrinkage and selection operator (LASSO) Regression is a technique that penalizes the absolute values of coefficients, effectively shrinking some coefficients to zero. All values with a coefficient bigger than zero are chosen as variables for the model.
LASSO Regression, initially devised as a linear regression model, addresses a prominent limitation, the influence of outliers, associated with traditional linear regression. In response to this weakness, Ridge Regression was introduced as an attempt to mitigate the  effects of outliers. 
Ridge Regression restricts the influence of certain variables by penalizing them and driving their coefficients closer to zero. Although Ridge Regression helps in stabilizing the model, it still necessitates the consideration of variables to some extent.
In a quest to further optimize model performance, including computational resources, LASSO Regression emerged as a valuable extension. LASSO Regression, distinct from Ridge Regression, employs a mechanism that shrinks the coefficients of unimportant variables to precisely zero, effectively excluding them from the model.
LASSO Regression introduces a form of variable selection by penalizing the absolute values of coefficients, ensuring that only variables with non-zero coefficients are retained and utilized in the model. This feature makes LASSO Regression particularly useful in cases where the identification of influential predictors is crucial, thereby enhancing the model's interpretability and efficiency \cite{muthukrishnanLASSOFeatureSelection2016}.

\subsubsection{Forward Feature Selection (FFS)} \label{chapter:forward-feature-selection-explain}
\begin{figure*}
    \centering
    \includegraphics[width=0.65\linewidth]{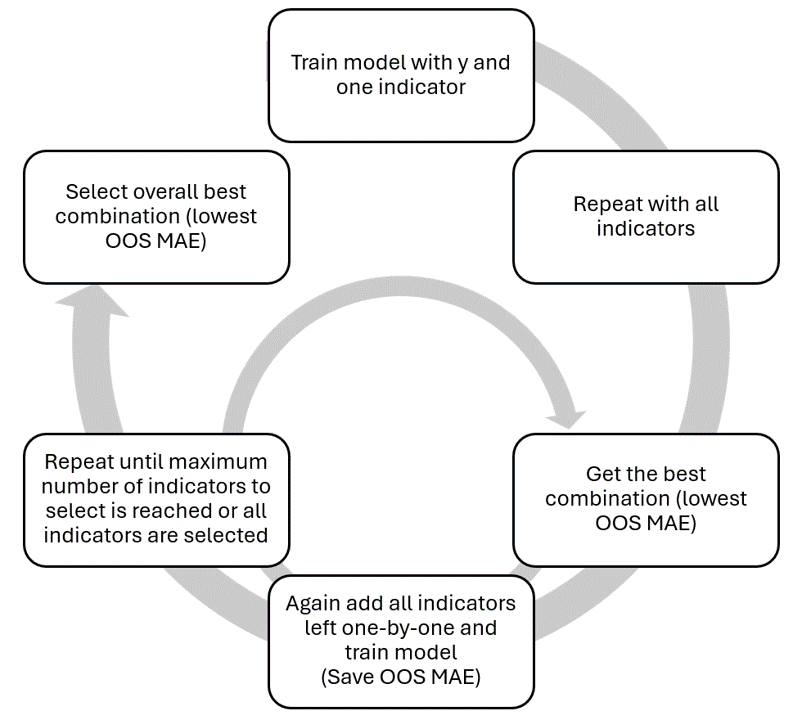}
    \caption{This figure shows an overview of the procedure applied by the implementation of the FFS in this work. OOS MAE stands for out-of-sample mean absolute error.}
    \label{fig:flow-chart-ffs}
\end{figure*}

Two further common feature selection strategies are forward and backward feature selection. These techniques are employed to refine the set of input features, thereby improving the model's performance. It is a systematic approach that starts with an empty set of variables. Subsequently, it adds one variable at each iteration, selecting the variable that provides the most substantial performance enhancement.
Backward Feature Selection commences with all available features and progressively eliminates those that do not contribute significantly to the model's performance, ultimately leading to a more streamlined feature set. However, it is computationally more demanding when dealing with many initially available features \cite{ververidisSequentialForwardFeature2005}.
The Forward Feature Selection was chosen for the experiment in this work. It will be performed with every 48 features for both models, saving the result of every feature combination to find the best of them. 
An overview of this method's procedure is given in \figref{fig:flow-chart-ffs}.

\subsubsection{Manual Selection} \label{chapter:manual-selection-explain}
The selection of exogenous variables for the forecasting model in this method is executed with consideration of the authors' domain expertise. This approach is based on the authors' intrinsic understanding of the subject.
The 48 datasets are viewed, and the most suitable datasets according to the domain knowledge of the authors are included in the forecast as exogenous variables. 
Furthermore, a direct benchmarking of manual selection based on domain knowledge against other selections, that seem similar, is unfeasible due to the subjective nature of domain expertise. 

\subsection{Experimental setup}
The Cross Industry Standard Process for Data Mining (CRISP-DM) separates the general data mining process into the steps of business understanding, data understanding, preparation, modelling and evaluation \cite{kotuDataScienceProcess2019}. The setup in this work, an individual adaption of the data science process to our research question and use case.

\textit{1. Data Pre-Processing: }
At this point in the process, the data is expected to be cleaned like explained in \secref{internal-ts}. This phase of the study involved preparation of pre-processed data suitable for both the SARIMAX and NP models. This process included merging market indicators with the demand data into a structured Dataframe using the \verb|pandas| library \cite{Pandas}. 

\textit{2. Hyperparameter Tuning for SARIMAX: }
Hyperparameter tuning was executed for the SARIMAX model. The aim was to identify optimal parameters automatically, primarily guided by the minimization of the out-of-sample Mean Absolute Error (MAE). It is important to mention again that this experiment focused on the comparison of variable selection methods rather than the performance of SARIMA under different parameter configurations. Consequently, the hyperparameter tuning resulted in an order of (62, 1, 4) and a seasonal order of $(0, 0, 0)_4$. So $p=62$ past values are differentiated $d=1$ times, incorporating $q=4$ past forecasting errors. $s=4$ indicates a quarterly seasonality. However, there is no seasonal order $P=D=Q=0$, meaning that the model does not include any seasonal autoregressive, seasonal differencing, or seasonal moving average components for the seasonal pattern. Therefore, no seasonality is modelled for this example.

\textit{3. Choose Forecasting Horizon: }
The forecasting horizon of 12 was chosen based on practical considerations, aligning with the standard practice in forecasting competitions. Additionally, it is a commonly used forecasting horizon in business context when planning the next fiscal year \cite{bojerKaggleForecastingCompetitions2021}.

\textit{4. Preparation of Training- and Test set: }
To facilitate rigorous testing and model evaluation, the dataset was partitioned into training and testing sets. The test set spanned 12 months 
and was used for testing the trained model.

\textit{5. Benchmarking Behavior: }
For benchmarking purposes, the behavior of both models was evaluated without the inclusion of any exogenous variables for each product. Also, two different training data time ranges are compared 2016.01-2021.04 (64 months) and 2019.01 -2021.04 (28 month). So six datasets are generated that will be named accordingly, e.g. BC1\_64 for demand data from product BC1 from 2016.01 until 2021.04 as training data. These two time ranges are chosen because it cuts the available training data in a half. 

\textit{6. Variable Selection with the different methods: }
Three automatic feature selection methods, namely correlation, LASSO regression, and the Forward Feature Selection, were employed to systematically identify influential variables. Additionally, the authors manually conducted a review of all 48 datasets. This review involved examining dataset titles and metadata from Eurostat to assess the relevance and appropriateness of the selected datasets. 

\textit{8. Models trained with and without the exogenous variables: }
Both SARIMAX and NP models were trained with the selected features for all six datasets to investigate their performance. 

\textit{9. Evaluation of the different variable selection methods: }
How the performance is evaluated is always also depending on the metric chosen to compare.
The error measure is the mean absolute error (MAE) (\eqnref{mae}) which records the mean difference between the forecast and the time series. 
The choice was made to assess the results of this work by examining the out-of-sample mean absolute error (OOS MAE). This decision was based on its simplicity and the need to select a specific metric for comparison and analysis at some stage.

\begin{equation} \label{mae} 
	MAE = \frac{1}{n} \sum_{t=1}^{n} |y_t - \hat{y}_t|
\end{equation}

During the test period, it is essential to acknowledge that the Ukraine war had a substantial impact on global and European markets, thereby potentially influencing the results and behavior of the forecasting models.

\section{Results and Discussion}
In this section, the results of the experiments are presented, and their findings are discussed.
First, the two models are compared regarding their performance without adding regressors between the two different training data time ranges chosen. Afterwards, the result for all selection methods are presented. The sections only present an in-depth analysis of the results for the BC1 time series and both training data ranges 64 months (2016.01-2021.04) referred to as BC1\_64 and 28 month (2019.01-2021.04) referred to as BC1\_28. The results for BC2 and BC3 should test results from BC1 and therefore mostly presented in tables but not with an in-depth analysis like BC1. This would extend this work's scope. 

\begin{table*}
	\centering
	\caption{Overview of all SARIMAX results for all datasets}
	\begin{tabularx}{\textwidth}{Xcccccc}
		\toprule
		Forecasting setting & BC1\_64 & BC1\_28 & BC2\_64 & BC2\_28 & BC3\_64 & BC3\_28   \\ 
		\midrule
		Without exogenous variables & 6.35 & 14.08 & 5.66 & 16.84 & 15.43 & 16.16   \\ 
		Correlation-Based Selection & 7.38 & 26.47 & 6.37 & 5.78 & 11.73 & 10.62 \\
	     \quad \textit{Nbr. Exogenous variables} & \textit{3} & \textit{4} & \textit{3} & \textit{4} & \textit{3} & \textit{4}    \\ 
		LASSO regression feature selection & 13.52 & 15.38 & 5.05 & 4.98 & 15.43 & 23.38 \\ 
		\quad \textit{Nbr. Exogenous variables} & \textit{3} & \textit{10} & \textit{9} & \textit{1} & \textit{0} & \textit{16}    \\ 
		Forward Feature Selection SARIMAX & \textbf{5.64} & \textbf{5.94} & \textbf{2.83} & \textbf{2.63} & \textbf{6.35} & \textbf{4.22} \\
		\quad \textit{Nbr. Exogenous variables} & \textit{1} & \textit{15} & \textit{13} & \textit{16} & \textit{26} & \textit{14}    \\   
		Manual feature selection & 13.99 & 18.11 & 7.37 & 8.73 & 10.67 & 26.97 \\
		\quad \textit{Nbr. Exogenous variables} & \textit{18} & \textit{18} & \textit{18} & \textit{18} & \textit{18} & \textit{18}    \\ 
		\bottomrule
	\end{tabularx}
	\caption*{Shows the OOS MAE to compare forecasting performance for the forecasting horizon of 12 month for the different variable selection methods tested on all datasets. FFS has the best results for both models and all datasets and training horizon lengths.}
	\label{tab:overview-all-results-SARIMAX}
\end{table*}

\begin{table*}
	\centering
	\caption{Overview of all NP results for all datasets}
	\begin{tabularx}{\textwidth}{Xcccccc}
		\toprule
		Forecasting setting & BC1\_64 & BC1\_28 & BC2\_64 & BC2\_28 & BC3\_64 & BC3\_28   \\ 
		\midrule
		Without exogenous variables & 11.14 & 13.65 & 3.75 & 9.08 & 17.53 & 9.06   \\
		Correlation-Based Selection & 14.11 & 11.6 & 4.78 & 7.02 & 19.28 & 11.24   \\
		\quad \textit{Nbr. Exogenous variables} & \textit{3} & \textit{4} & \textit{3} & \textit{4} & \textit{3} & \textit{4} \\
		LASSO regression feature selection & 14.55 & 13.6 & 5.73 & 6.29 & 17.31 & 28.79 \\
		\quad \textit{Nbr. Exogenous variables} & \textit{3} & \textit{10} & \textit{9} & \textit{1} & \textit{0} & \textit{16}   \\ 
		Forward Feature Selection NP & \textbf{9.16} & \textbf{9.09} & \textbf{2.9} & \textbf{3.36} & \textbf{8.14} & \textbf{8.49} \\   
		\quad \textit{Nbr. Exogenous variables} & \textit{28} & \textit{14} & \textit{17} & \textit{8} & \textit{14} & \textit{14}    \\ 
		Manual feature selection & 10.13 & 16.1 & 2.97 & 16.07 & 28.3 & 18.23  \\
		\quad \textit{Nbr. Exogenous variables} & \textit{18} & \textit{18} & \textit{18} & \textit{18} & \textit{18} & \textit{18} \\
		\bottomrule
	\end{tabularx}
	\caption*{Shows the OOS MAE to compare forecasting performance for the forecasting horizon of 12 month for the different variable selection methods tested on all datasets. FFS has the best results for both models and all datasets and training horizon lengths.}
	\label{tab:overview-all-results-NP}
\end{table*}

\subsection{Comparison of SARIMAX and NeuralProphet model}
\begin{figure*}
	\centering
	\includegraphics[width=1\linewidth]{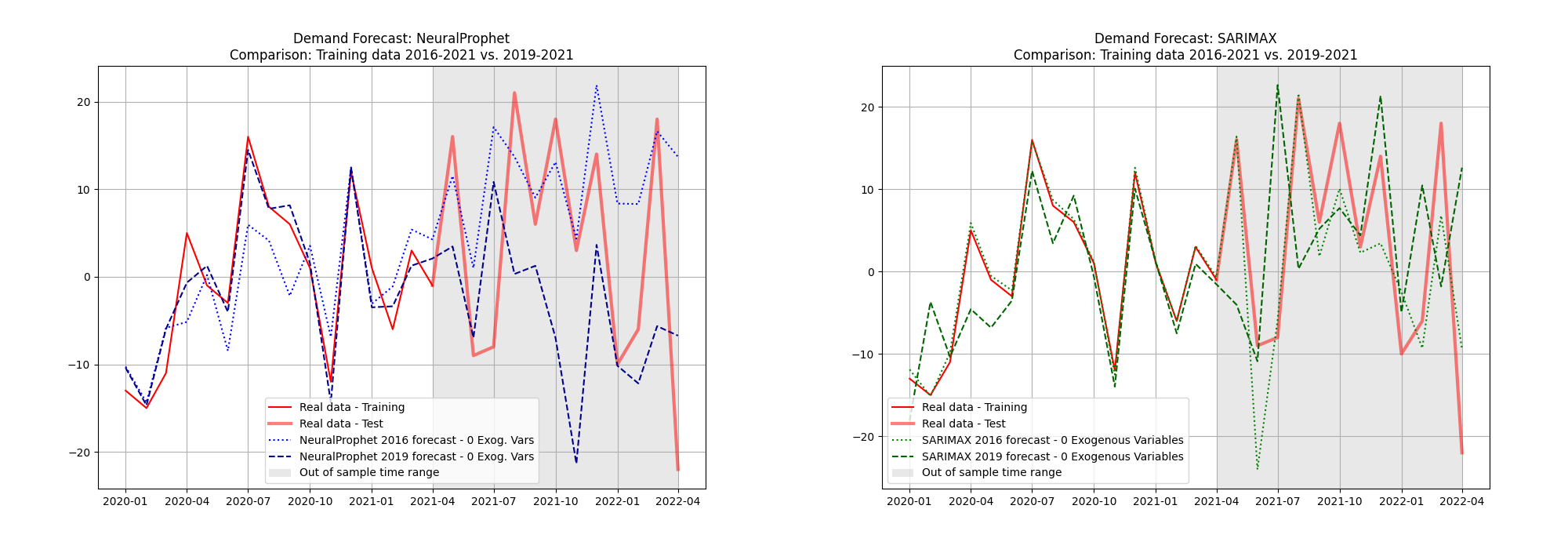}
	\captionsetup{font=small}
	\caption{This figure shows the \textbf{NP model} (left) and \textbf{SARIMAX model} performing on the different time ranges of normalized training data (BC1\_64 (2016) and BC1\_28 (2019)) without adding exogenous variables. This plot only shows the time range from 2020 on to have better visibility on the out-of-sample behavior of the forecasts (2021.05-2022.04).}
	\label{fig:forecast20162019}
\end{figure*}

First, the behavior of the models is discussed on the demand time series from cleaning product BC1 without adding any regressors. This will contribute to the answer for RQ3. This section is taking a look at the different behavior of the two models for the two different training data time ranges. Differences and similarity in performance and behavior are analyzed. Details on how the forecasts from the models differ when adding an exogenous variable to the model can be found in \secref{chapter:sarimax} and \secref{chapter:np}. It is also worth noting that the runtime of the different approaches, despite Forward Feature Selection, does not differ that much (between 13 seconds and five minutes).
The performance of the SARIMAX model trained on the two different time intervals without the inclusion of regressors showed the following results:
The OOS MAE is significantly better for the longer time range for BC1\_64, in comparison to the shorter time span for BC1\_28 (6.35 versus 14.08). The same holds true for BC2, but OOS MAE BC3\_64 is only slightly worse than BC3\_28 (15.43 vs. 16.16) (see \tabref{tab:overview-all-results-SARIMAX}). To prove this behavior, more products have to be considered. 
For BC1, the distinction becomes clear upon examining \figref{fig:forecast20162019}. Here, the forecast for the longer forecasting horizon shows greater accuracy and closely aligns with the real data from this extended time frame.
Of particular note is the discernible difference in trend behavior at the end of the forecast: in the longer time range, the forecast trend follows the same direction as the test and real data, whereas the forecast derived from the shorter training data period shows a contrasting trend.
These observations underscore the sensitivity of SARIMAX model performance to the choice of time interval and how predictions can diverge for different data segments.
For the NP model, also interesting observations emerge when comparing its performance on two distinct training data time ranges. The OOS MAE does not show a substantial difference between the two training data time ranges for BC1 (BC1\_64 vs. BC1\_28) but not for BC2 and BC3 (see \tabref{tab:overview-all-results-NP}). For BC1\_64, the OOS MAE is 11.14, while for BC1\_28, the OOS MAE is 13.65. This disparity in OOS MAE values can be elucidated by the fact that the NP model demonstrates a reasonably close match between its predictions and the actual data, except a noticeable dip in November 2021 for the model trained on BC1\_28. It is worth noting that NP autonomously determines seasonality without the need for manual hyperparameter tuning, which can be advantageous for practical use in real-world scenarios, making it a valuable tool for end users in a business context, but it probably also can detect false seasonalities. Interestingly for BC3 the OOS MAE for the shorter training data time range BC3\_28 is better than the longer one BC3\_64 (17.53 vs. 9.06) (see \tabref{tab:overview-all-results-NP}). The three different products viewed for two different training data time ranges do not provide representative information on whether NP generally performs better on short or long term training data time ranges.
The dip which can be seen in the forecast made from BC1\_28 in \figref{fig:forecast20162019} in November 2021 probably occurred because of the dip in 2020 and the one year before. As NP automatic hyperparameter tuning was used, the model probably detects a yearly seasonality here. This leads to the assumption that NP also performs more reasonable with a longer training time range.
From the economic context and the crises we are facing the last years, it is known, that these dips are not caused by seasonality. The extreme dip at the end of the test data set can also be attributed to the economic situation as impact of the start of the Ukraine war in February 2022 \cite{Allam2022}. Consequently, this event is considered an outlier in the time series.
From this perspective, it is intriguing to observe that the NP model's trend at the end of the forecast consistently depicts a downward direction, as depicted in \figref{fig:forecast20162019}. This trend remains consistent regardless of whether the forecast is generated by the model trained on the shorter or longer training data time range.
This final trend at the end of the training data and the major dip in the forecast of the 2019-2021 trained model in the forecast for 2021.11 are the main differences we can see here between SARIMAX and NP model when adding no exogenous variables to the model.

To contribute to RQ3, this is how the forecasting models differ for the example time series of product BC1 without adding exogenous variables:
Looking at all results in \tabref{tab:overview-all-results-SARIMAX} and \tabref{tab:overview-all-results-NP} it can be observed that the performance of both models depends on the time series that should be forecasted and on the market indicator time series that are taken into account.
However, it is essential to recognize that achieving a close fit like that of SARIMAX for BC1\_64 may occur randomly. More importantly, a forecasting model should prioritize generalization to ensure the robustness of forecasts. Consequently, the strong trend of the  SARIMAX model towards the end of the forecasting horizon works effectively for BC1\_64 but not for a shorter training time series of BC1\_28, where the trend moves in the opposite direction of the test data. In comparison, both NP models (BC1\_64 and BC1\_28) appear to show greater generalization, as they reflect the correct trend at the end of the forecasting horizon. Further differences in performance of the models, taking into account the results of the analysis of methods of feature selection, is conducted in \secref{chapter:comparison-of-fs-methods}. 

\subsection{Analysis of Methods of Feature Selection}
As stated in \secref{chapter:contribution} with RQ1, the main question for this work was, if and how a forecast can be optimized by adding market indicator time series to a forecasting model. In the following subsections, the results of the exploratory approach to test common feature selection techniques, are presented and analyzed. 
Within the subsections, the difference in performance and behavior of a model itself with different training data time ranges and the difference of the models are discussed for each of the feature selection approaches.
For every approach, an example is given for what datasets from Eurostat were chosen.

\subsubsection{Correlation-Based Selection}
For BC1\_64 the three variables derived are an indicator showing the monthly crude oil supply \cite{MonthlyCruideOil}, the monthly data of crude oil imports by fields of production \cite{CrudeOilImports} and the monthly development of import prices in industry in general \cite{ImportPricesIndustry}. It is quite interesting to see that two time series containing information about the crude oil supply for Germany are correlating with the demand of the company offering the real world data.
The four variables chosen because of their correlation to BC1\_28 do not have an intersection with the three correlating the most with BC1\_64, which underlines the conclusion that the structure and course of BC1\_64 and BC1\_28 are different from each other. First, there is a food pricing monitoring indicator \cite{FoodPriceMonitoring, MetadataFoodPrice}, which is surprising because it would not be an intuitive time series to incorporate as an exogenous variable to support / optimize a forecast. The second correlating time series shows the long term development of an interest rate of the European Monetary Union 
\cite{EMUConvergenceCriterion, MetadataMaastrichtCriterion}. Also, a retail sales indicator was chosen to be integrated to the model by correlation analysis \cite{RetailSaleMonthly, MetadataBusinessConsumer}, which is as unexpected to incorporate as the food pricing monitoring because the industrial research partner is mostly acting in the business to business (B2B) market.
Lastly also a short-term interest rate time series was found to be added to the model \cite{InterestRatesMonthly, MetadataMonetaryFinancial}.
The inclusion of market indicators from the correlation analysis did not significantly optimize the forecasting model performance of SARIMAX. The impact of adding exogenous variables varies depending on the training data time range and the dataset.
For the BC1\_64, the OOS MAE remained relatively stable, with a minor increase from 6.35 without exogenous variables to 7.35 with their inclusion. In contrast, for the shorter training data range of BC1\_28, the addition of exogenous variables led to a substantial change in OOS MAE, soaring from 14.08 without variables to 23.69 with their incorporation. A detailed summary of the OOS MAE results can be found in \tabref{tab:overview-all-results-SARIMAX}. For three of the six datasets the resulting OOS MAE is better than the one from the forecast made without exogenous variables in the model (BC2\_28, BC3\_64, BC3\_28). It is reasonable to assume that this is not a random occurrence. To ascertain that this correlation is not coincidental and to identify the specific conditions that have caused the correlated variables to provide enhanced support for the forecast, more comprehensive experiments need to be conducted.
For BC1\_28 it optimizes the forecast from OOS MAE without variables 13.65 vs. OOS MAE of 11.60 with the variables. As stated in \secref{chapter:np}.
the BC1\_28 forecast without exogenous variables is not forecasting the test data well (e.g. having a huge dip at 2021.11). This finding can lead to the assumption that for shorter training data time ranges, it is advantageous to integrate market indicators and with that more context to the forecasting model. To validate this assertion, additional experiments with diverse datasets from various industries and across multiple training data time ranges must be conducted. When adding the exogenous variables found with the derived correlation analysis to the NP model trained on BC1\_64 the forecast nearly does not change. 
The variables that correlated to BC1\_64 and BC1\_28 are an interest rate time series and time series about consumer goods. Those would not be very likely added intuitively because the industrial research partner mostly sells its industrial cleaning supplies and machines to B2B clients.
The time series chosen by correlation based selection supported both model's forecast partly. To address RQ3, optimization was achieved for two out of the three NP models trained with a shorter training date range by introducing correlated variables. Additionally, SARIMAX forecasts were optimized for three out of the six datasets. This suggests that using the correlation-based selection method can improve the forecast. To proof that this is not a random occurrence and to further understand the conditions under which correlated variables enhance a forecasting model, additional experiments need to be conducted.

\subsubsection{LASSO Regression}
For both BC1\_64 and BC1\_28, LASSO regression predominantly identifies the same indicators from Eurostat. For instance, in both cases, LASSO selects an indicator related to the financial balance of an economy, encompassing all transactions associated with changes in ownership of foreign financial assets and liabilities \cite{MetadataBalancePayments, FinancialAccountMonthly}. Another example is a time series representing the monthly construction volume of a country \cite{ConstructionMonthlyDatab, MetadataIndustryTrade}. An example of intersection with other variable selection is the inclusion of a retail trade indicator, indicating the turnover growth rate for retail sales. This variable is also chosen by the forward selection using the NP model for dataset BC1\_64 and forward selection using the SARIMAX model for dataset BC1\_28 \cite{RetailTradeMonthly, MetadataIndustryTrade}.
In the case of the NP model for BC1\_28, LASSO regression optimization resulted in only a minor improvement when compared to the model without exogenous variables (OOS MAE of 13.60 with LASSO optimization versus 13.65 without exogenous variables, find this results in \tabref{tab:overview-all-results-NP}). Also, for BC2\_28 there was a small improvement in OOS MAE compared with the forecast without exogenous variables (see \tabref{tab:overview-all-results-NP}). But as the improvements are not significant, they can be assumed to be achieved randomly.
Similarly, for the SARIMAX model, the introduction of LASSO regression optimization had a slight adverse effect on performance, specifically for the shorter training data interval of BC1\_28 (OOS MAE of 15.34 with LASSO optimization versus 14.08 without exogenous variables). Interestingly, forecasting OOS MAE was improved by using exogenous variables identified with LASSO for BC2 (see \tabref{tab:overview-all-results-SARIMAX}). But as it is only the case for two of the six datasets, it may be considered a random occurrence.
From the higher OOS MAE (14.55 with vs. 11.14 without variables) (see \tabref{tab:overview-all-results-NP}) it can be deducted that the selected variables did not improve the forecast. But for BC1\_28 the 10 market indicators added from conducting a LASSO regression feature selection improve the results a little as stated at the beginning of this section.  
The slight optimization from 13.65 OOS MAE to 13.60 can be seen in \tabref{tab:overview-all-results-NP} but is not significant and cannot be proofed from the results from the other datasets.
Reconsidering the research questions RQ1 and RQ2, we can learn from this experiment that selecting the market indicators to incorporate to a forecast with LASSO regression is not beneficial for the forecasting performance of both forecasting methods. The observation, that the LASSO method's selected variables match with some variables selected by more effective variable selection techniques, lead to the assumption that also the optimal combination of the market indicators to incorporate is important.
\subsubsection{Forward Feature Selection}
\begin{figure*}
    \centering
    \includegraphics[width=1\linewidth]{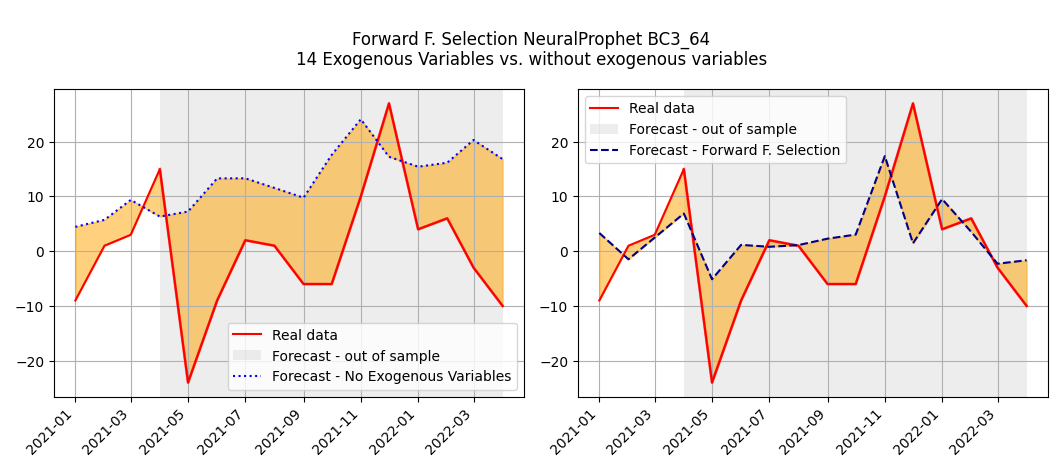}
    \captionsetup{font=small}
    \caption{This figure shows NP model's forecast when adding the exogenous variables found with the Forward Feature selection using NP model vs. the forecast of the model without exogenous variables for \textbf{BC3\_64}. This plot only shows normalized data for the time range from 2021 on to have better visibility on the out-of-sample behavior of the forecasts (2021.05-2022.04).}
    \label{fig:forecastnpfs2019}
\end{figure*}

The variables selected by the Forward Feature Selection varied between the model used for the forward selection and the different datasets.
As stated in \secref{chapter:sarimax} the forward selection using SARIMAX on dataset BC1\_64 selects only one variable as the best combination. It is a monthly business indicator for services  \cite{ServicesMonthlyData, MetadataBusinessConsumer}
Instead for BC1\_28 15 features were selected. 
Another example is a retail trade indicator which was also selected by LASSO regression as well as indicators related to the financial balance of an economy  \cite{FinancialAccountMonthly, MetadataIndustryTrade, RetailTradeMonthly}.
NP found 28 different variables as the best combination for forecasting dataset BC1\_64 and 14 for BC1\_28. Both of them also selected a retail sales indicator \cite{RetailSaleMonthly, MetadataBusinessConsumer}. As retail sales seem not connected to the sale of B2B cleaning supplies, this shows how a market indicator may support the forecast that would not be considered intuitively.
Another related indicator showing the turnover volume of sales was also selected for both training data time ranges \cite{TurnoverVolumeSales, MetadataShorttermBusiness}.
Two different indicators on producer prices were chosen for BC1\_28 \cite{ProducerPricesIndustry, MetadataShorttermBusiness} and for BC1\_64 \cite{MetadataShorttermBusiness, ProducerPricesIndustry}. Producer prices seem quite close to the economical context of our industrial research partner. For BC1\_64 a monthly energy consumption indicator was selected into the best working combination as well \cite{EnergyMonthlyData, MetadataIndustryTrade}.
Another variable added to the model is "Arrivals at tourist accommodation establishments" which is interesting as it seems to have nothing in common or no influence on the demand of industrial cleaning supplies \cite{ArrivalsTouristAccommodation, MetadataOccupancyTourist}.
The inclusion of exogenous variables introduced a notable enhancement to the forecast in comparison to models without these regressors, demonstrating improved performance across all six datasets and forecasting models (see \tabref{tab:overview-all-results-SARIMAX}, \tabref{tab:overview-all-results-NP})
For BC1\_64, a single market indicator was added for the SARIMAX model, which contributed to optimizing model performance, achieving the best results.
This improvement is evident in the OOS MAE, where both training data ranges, 2016 and 2019, witnessed substantial enhancements in forecast accuracy (BC1\_64: 6.35 without exogenous variables versus 5.64 with exogenous variables, as presented in \tabref{tab:overview-all-results-SARIMAX}).
These findings emphasize the significance of the Forward Feature Selection approach in augmenting SARIMAX model performance through the incorporation of exogenous variables. The resulting forecasts are visibly closer to the actual data, underscoring the effectiveness of this variable selection strategy.
Also, for NP, there was a significant optimization in the OOS MAE in comparison to the basic model performance. OSS MAE of BC1\_64 was at 11.14 without the variables and landed at 9.16 by adding the variables to the forecasting model during training. For the BC1\_28 the impact was even higher: OOS MAE 13.65 without and 9.09 with exogenous variables (see \tabref{tab:overview-all-results-NP}). Especially until 2022.11, which is 7 months into the forecasting horizon, the NP forecast is extremely close to the test data. But the fact that it for example does not get the trend right at the end of the horizon, but the model without variables does, leads to the assumption that adding the variables in this setting may help more when having shorter forecasting horizons like 6 months for example. It is also interesting to see in \figref{fig:forecastnpfs2019} that the forecast of NP when adding the exogenous variables improves into the direction of the real values and gets the trend at the end of the horizon slightly better, but still does not look like a good fit. But this is also proofed with the overall worse OOS MAE of NP than SARIMAX.

\begin{figure*}
    \centering
    \includegraphics[width=0.8\linewidth]{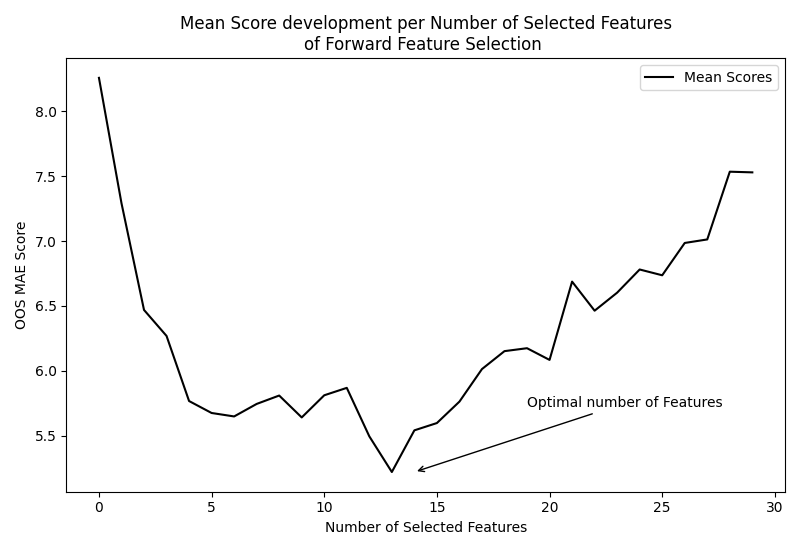}
    \caption{This figure shows the \textbf{mean OOS MAE score development} per number of selected variables for the Forward Feature Selection for all the Forward Feature Selection experiments discussed in this work.}
    \label{fig:means_score_dev}
\end{figure*} 

The Forward Feature Selection methodology implemented in this study iteratively evaluates all potential predictor variables, as delineated in \secref{chapter:forward-feature-selection-explain}. Subsequently, the optimal variable subset is selected. A more computationally efficient approach would entail establishing a predefined limit on the number of variables to select. But initially, it was unclear what could be the ideal number of variables to select. To identify this for future experiments, the OOS MAE for every number of variable selected up to the total of available market indicators (48) was recorded for the six time series analyzed. \figref{fig:means_score_dev} depicts the trend of the mean OOS MAE over the number of selected variables from all experiments. A discernible increase in the mean OOS MAE beyond the inclusion of 13 variables suggests that future experiments could cap the selection at this threshold to enhance computational efficiency. To account for variability and ensure robustness, a margin could be applied, extending the limit to 20 variables. A hard boundary will optimize the runtime as a lot less variable combination need to be tested. It is imperative to note that this recommendation is predicated on the findings derived from the six time series included in the present study. Theoretically, it could also be interesting to see if the OOS MAE decreases again if a lot more variables selected by the method will be incorporated. As the computational demand of this is high, it is not interesting for the practical use of this method, e.g. in a company setting.
It is also worth noting that this feature selection technique is dependent on the model used for selecting the feature and the error measure used. So on the one hand good results can be expected, but on the other hand there is a risk for over fitting. Choosing an out-of-sample error measure (OOS MAE) to optimize was supposed to avoid this problem. It means the model was trained on one of the training date ranges (2016.04-2021.04 or 2019.04.-2021.04). Then with this model a forecast was made for the test horizon (2021.05-2022.04) for data the model does not know. This out-of-sample forecast was compared with the real test data from the test horizon.
The results of this experiment contribute to RQ1 and RQ2 in the following manner: The use of the Forward Feature Selection as a method for selecting external variables led to optimization of the OOS MAE for all datasets, encompassing three different products and both training data time ranges. These consistent improvements suggest that Forward Feature Selection is likely to yield enhanced forecasting results, and this hypothesis should be further validated through additional experiments with different datasets.

\subsubsection{Manual Selection}
The specialty of the manual variable selection by the authors is that it is subjective and that it is the same number of variables added for every model and both training date range. It was tried to simulate the process of a practitioner watching a list of possible variables to incorporate and intuitively sort the ones out that do not seem relevant (as explained in \secref{chapter:manual-selection-explain}). A good example for this is the data showing "Arrivals at tourist accommodation establishments" which is added to the model by the automatic approach of the Forward Feature Selection but was not considered when selecting the exogenous variables manually.
An illustrative indicator that appears to support the forecasting task is constructed based on a monthly survey conducted among a representative selection of industrial companies. This survey covers various aspects, including their production levels, order book status, inventory of finished products, perceptions of economic uncertainty, selling prices, and employment data \cite{IndustryMonthlyData, MetadataBusinessConsumer}. Other examples are various producer price indicators \cite{MetadataShorttermBusiness, ProducerPricesIndustry, ProducerPricesIndustrya, ProducerPricesIndustryb}.
The manual variables selection optimizes the result for the OOS MAE  only at random occasions. For SARIMAX the forecast for BC2\_28 and the one for BC3\_64 are optimized. For NP, the OOS MAE is better for BC1\_64 and BC2\_64.
For the one experimental set up where the manual selection optimized the result of the forecast, it can be observed that the forecast made by the model with the 18 variables added indeed did not differ much from the one without variables. This leads to the hypothesis that some of these 18 variables overlay each other so that it results in close to zero effect. So the forecast is slightly better in this special case (BC1\_64 from the industrial research partner) but the selected variables do not change the forecast. So it can be assumed that it is just a slightly better forecast by random occasion.
The manual selection approach is inherently subjective and lacks the objectivity required for a scientific comparison with other selection methods. Replicating this experiment precisely on a different dataset is unfeasible due to its subjective nature. It is included in this context to replicate the intuitive decision-making process that a practitioner might follow when identifying time series variables with the potential to enhance the forecast.
Overall, Manual Selection shows the worst results, pointing out selecting indicators only based on the knowledge of human experts does not necessarily lead to optimized forecasting results.

\subsection{Comparison of Methods of Feature Selection} \label{chapter:comparison-of-fs-methods}
\begin{figure*}
    \centering
    \captionsetup{font=small}
    \includegraphics[width=1\linewidth]{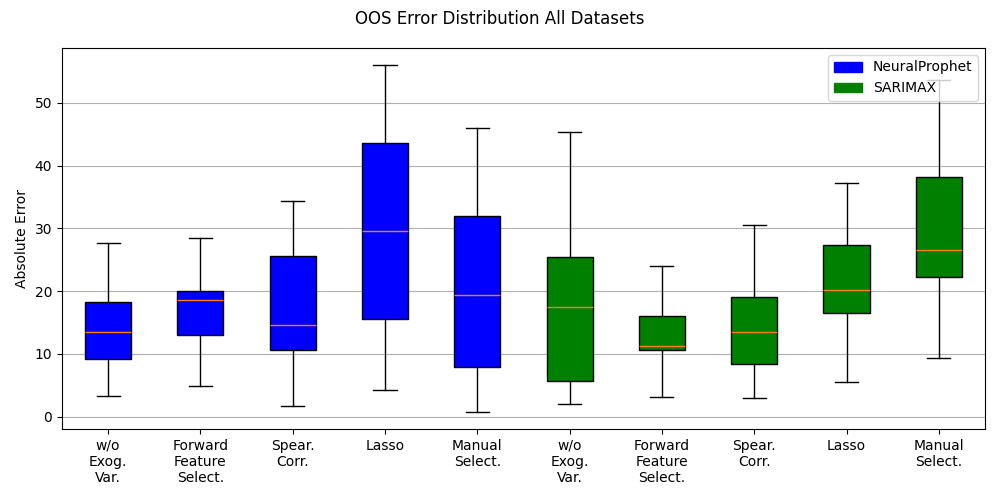}
    \caption{This box plot shows the absolute error distribution without outliers for all variable selection methods tested in this work's experiments for both methods and across all datasets.}
    \label{fig:boxplot-all-errors}
\end{figure*}

It needs to be stated that these variable selection methods are only tested on three products for two different time ranges resulting in six different datasets, because the focus of this work is the explicit forecasting task of the research project partner company testing all those approaches. For verifying and generating more general statements about the different variable selection methods, more tests on data from other scenarios (like different industries) need to be considered. The box plot in \figref{fig:boxplot-all-errors} shows that, looking at the median (orange line in the middle of the boxes), for NP the Forward Feature Selection has a higher median than the first box showing the forecasting error without exogenous variables. As the mean error is lower for the Forward Feature Selection compared to the version without indicators, it can be concluded that the error distribution for the Forward Feature Selection with NP is left-skewed. This indicates that there are a few very low errors that reduce the mean, indicating that for NP, the Forward Feature Selection does not perform as well as the OOS MAE indicates. This should be looked at closely in further experiments. For SARIMAX the box plot (\figref{fig:boxplot-all-errors}) shows clearly that Forward Feature Selection works better and has a smaller error distribution than without exogenous variables. The same can be observed for the correlation-based method. 
The box plot in \figref{fig:boxplot-mae} illustrates that overall the models do not differ much in performance as the boxes are all on the same level for every method. This was also shown by a significance test. The OOS MAE results of the two models do not differ significantly. This leads to the conclusion that the forecasting model is less important than the selected variables to the forecasting error. But looking at \figref{fig:boxplot-all-errors} as stated above, it can be seen, that the Forward Feature Selection does not work stable for NP, looking at the periodic error distribution. Further research needs to take a closer look at this and what could be the reason. But, the error distribution in \figref{fig:boxplot-all-errors} shows the higher variance in errors of the forecast without adding exogenous variables. Therefore, a recommendation of action at this stage is to use NP for forecasting without exogenous variables and SARIMAX combined with the Forward Feature Selection when adding exogenous variables.

Choosing only Eurostat data was a decision to ease the task of finding the right indicators in the first place. It can be assumed that taking into account more variables from different sources will improve the results further. It is a task for future research to identify different information and time series from other sources.

The following core findings can be concluded in terms of RQ1 and RQ2 when comparing all feature selection methods experimental results (see \tabref{tab:overview-all-results-SARIMAX}, \tabref{tab:overview-all-results-NP}, \figref{fig:boxplot-all-errors}, \figref{fig:boxplot-mae}):

\begin{figure*}
    \centering
    \includegraphics[width=1\linewidth]{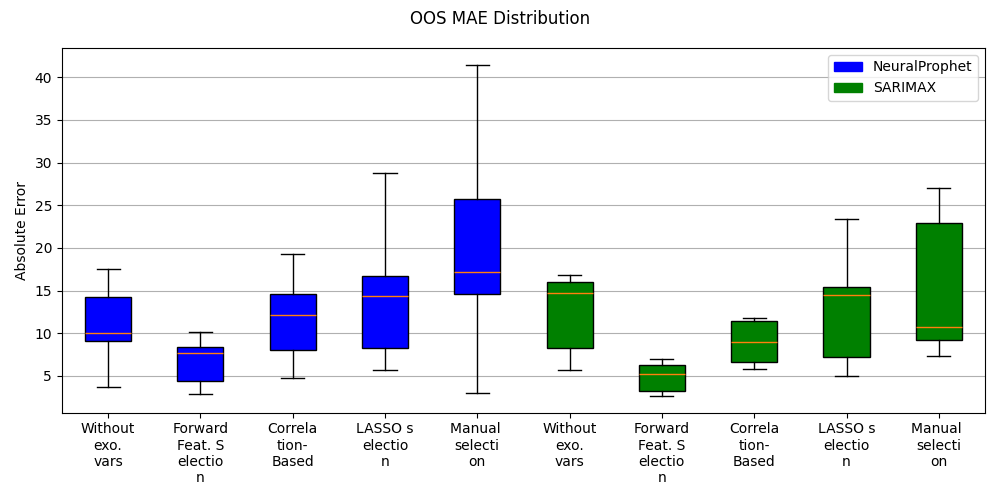}
    \captionsetup{font=small}
    \caption{This box plot shows the OOS MAE per dataset grouped by the selection methods.}
    \label{fig:boxplot-mae}
\end{figure*}

\textbf{Correlation-based variable selection enhances forecasts for shorter training data time range}: It was proven that correlation-based selection improves the forecasts of both models for shorter training data time ranges. Referring to RQ1 and RQ2 variables found with correlation analysis can support time series forecasting for shorter training data time ranges. This suggests the potential advantages of including market indicators in forecasting models found with correlation analysis. It also raises the hypothesis that shorter training data intervals benefit from the incorporation of external variables, as they provide valuable context and data.
To validate this hypothesis, further experiments with diverse datasets from various industries and several different training data time ranges are essential. 

\textbf{Exploring unconventional time series variables}: From all three automatic variable selection methods (Correlation, LASSO, Forward Feature Selection) it can be concluded that the introduction of unconventional time series variables underscores the value of experimentation in adding diverse variables to forecasting models. It highlights the inherent uncertainty regarding which variables may be relevant and beneficial for the forecasting task. This underscores the importance of exploring different variables to enhance model performance. The consequence of this finding is to perform variable testing with a broad set of variables, that would not be chosen intuitively, instead of manually pre-selecting potential variables too early in the process. 

\textbf{LASSO selection does not improve the forecast}: The enhancements it introduces to SARIMAX are applicable to only one product, which could potentially be a random occurrence. Similarly, for NP, the improvements in forecasting performance for shorter training data durations are not significant. 

\textbf{Forward Feature Selection has significantly better results compared to the models without exogenous variables}: Notably, the best feature selection technique tested in this work's experiments is the Forward Feature Selection. This was demonstrated by performance optimization across all six datasets, which can also be observed in \figref{fig:boxplot-mae}. 
Finding the right combination of the market indicators to incorporate is assumed to be more important for the result than a single exogenous variable. As the Forward Feature Selection does test a lot of different combinations, it is reasonable that this is the best approach. However, it is imperative to acknowledge that this approach is computationally intensive. For instance, conducting the Forward Feature Selection for both models with the BC1\_64 required approximately 12 to 15 hours, while the same process for the BC1\_28 took 3 to 4 hours. The experiments were conducted on a standard computer with the following specifications: Processor - AMD Ryzen 5 5600G with Radeon Graphics, 3.90 GHz, 16.0 GB RAM.
Given the computational demands, optimizations within the code and access to additional computational resources are needed to accelerate this process. As the forecasting horizon typically extends up to one year, even such a long calculation time could be acceptable for business practice. An additional periodic update of such models could also be considered. But if the forecasts are more time critical, the solution of the Forward Feature Selection in this setting is not ideal. Note that when taking into account more than 48 market indicators, this runtime growths as the model needs to be fit with every available exogenous variable to find the one leading to the lowest error or best fit. But the features probably do not need to be re-selected as often as a new forecast is needed, then such a long runtime for the selection would be practicable. Running both models with the exogenous variables is quite fast (between 11 and 30 seconds).
This also leads to the conclusion that the Forward Feature Selection approach can be used for meaningful variable selection with appropriate use of computational resources and time. RQ1 and RQ2, asking how a forecast can be optimized by adding external information and how to find them, can be answered accordingly with the Forward Feature Selection method. 

\subsection{Limitations} \label{chapter:limitations}
When conducting the experiments, certain aspects were not taken into account, which represent the limitations of this study:
\begin{itemize}[leftmargin=*]
	\item \textit{Need for comprehensive validation}: To affirm the observed improvements, it is essential to conduct further experiments with diverse datasets originating from various industries and encompassing a broader range of training data time intervals. This validation is crucial to determine the generalizability of the findings to different contexts and domains.
	\item \textit{Different forecasting horizons}: To ease the experiment task different training data time ranges (64 and 28 month) are tested but only one forecasting horizon of 12 months. Shorter and longer forecasting horizons (e.g. 3, 6 or 24 months) needs to be tested too.
	\item \textit{Confidence intervals}: For real-world applications, the inclusion of prediction or confidence intervals is common. However, determining the appropriate range or confidence level for these intervals presents a challenge that needs further investigation.
	\item \textit{Addressing variable redundancy}: It is possible that certain market indicators may overlap or offset each other's effects, resulting in a net impact close to zero. A solution might involve automated sorting or the exclusion of highly similar variables to optimize resource utilization.
	\item \textit{Examination of bigger possible market indicator dataset}: The consideration of a substantial dataset e.g. containing over 11,000 time series presents a significant challenge in terms of data management, analysis, and interpretation, requiring dedicated exploration.
\end{itemize}

\section{Conclusion and future research}
This work utilized SARIMAX and NP forecasting models to investigate, how real-world sales predictions can be improved by adding market indicator time series as exogenous variables from the Eurostat database (RQ1). Additionally, it examined the automated selection of appropriate exogenous variables through established feature selection methods (RQ2). Finally, a comparative benchmarking of SARIMAX and NP was conducted to evaluate their performance considering exogenous variables (RQ3). 

With regard to RQ1 and RQ2, the forecasts could be significantly improved by incorporating exogenous variables using both NP and SARIMAX. Particularly, the Forward Feature Selection method emerged as highly effective for automatically selecting and integrating features into the forecasting procedures. However, other tested methods such as LASSO have worse results than the forecast without adding indicators. Perhaps the most unexpected finding is, that the manual selection of external time series based on human expert knowledge yields the poorest forecasts in our experiments.

Looking at RQ3, the performance of SARIMAX and NP in terms of MAE OOS does not show significant divergence. This suggests that finding suitable market indicators is more important than selecting the base forecast model. Nevertheless, SARIMAX has a more stable performance when adding exogenous variables. Based on the experiments conducted, we recommend using NP without exogenous variables and SARIMAX with the Forward Feature Selection when integrating exogenous variables. 

Summarized from a practical standpoint, automating the selection and integration of external time series into forecasting models holds significant promise for enhancing sales and logistical operations. However, considering the limitations of this work (see Section \ref{chapter:limitations}), the following future research needs can be derived:

\begin{itemize}[leftmargin=*]
    \item \textit{Generalize findings}: Given that this study's experiments were limited to six real-world datasets, the results of this work need to be validated with more base time series and more exogenous time series to select from to generalize the findings.
    \item \textit{Runtime optimization}: To make the Forward Feature Selection more interesting for forecasts that need to be available fast, the runtime of the method need to be improved significantly. Using parallelization, feature reduction or faster forecasting approaches can help here.
	\item \textit{Exploration of additional data sources}: While the study focuses on Eurostat data, the integration of data from other sources, especially from major exporting countries, should be considered to enhance the accuracy and relevance of forecasts.
	\item \textit{Automation of data source discovery}: Developing a model that autonomously identifies relevant data sources based on provided topic information is an area for potential research, streamlining the data acquisition process.
	\item \textit{Consideration of neural network architectures}: While this study prioritizes interpretability by employing relatively simpler models, it is acknowledged that neural network architectures hold the potential to improve forecasting accuracy. Future research may explore the integration of neural networks for enhanced performance.
	\item \textit{Hyper parameter tuning for neural prophet}: Enhancing the performance of the NP model through hyperparameter tuning is a promising avenue for further research, seeking to maximize its predictive capabilities.
\end{itemize}

\section*{Author contributions (CRediT)}
\textbf{Lina Döring}: conceptualization, methodology, formal analysis, investigation, 
writing—original draft, writing—review and editing;
 
\textbf{Felix Grumbach}: data  curation, writing—review and editing, supervision
 
\textbf{Pascal Reusch}: project administration, funding acquisition, resources

\bibliographystyle{unsrt}
\bibliography{references}  

\begin{thebibliography}{10}

\bibitem{bennettAutoregressiveExogenousVariables2014}
Christopher Bennett, Rodney Stewart, and Junwei Lu.
\newblock Autoregressive with {{Exogenous Variables}} and {{Neural Network Short-Term Load Forecast Models}} for {{Residential Low Voltage Distribution Networks}}.
\newblock {\em Energies}, 7(5):2938--2960, April 2014.
\newblock Available online at \url{http://www.mdpi.com/1996-1073/7/5/2938}, Accessed: October 2023.

\bibitem{yerlikayaClimateChangeForecasting2020}
Bayram~Ali Yerlikaya, Seher {\"O}mezli, and Nazl{\i}can Aydo{\u g}an.
\newblock Climate {{Change Forecasting}} and {{Modeling}} for the {{Year}} of 2050.
\newblock In Shah Fahad, Mirza Hasanuzzaman, Mukhtar Alam, Hidayat Ullah, Muhammad Saeed, Imtiaz Ali~Khan, and Muhammad Adnan, editors, {\em Environment, {{Climate}}, {{Plant}} and {{Vegetation Growth}}}, pages 109--122. {Springer International Publishing}, {Cham}, 2020.
\newblock Available online at \url{http://link.springer.com/10.1007/978-3-030-49732-3\_5}, Accessed: October 2023.

\bibitem{ensafiTimeseriesForecastingSeasonal2022}
Yasaman Ensafi, Saman~Hassanzadeh Amin, Guoqing Zhang, and Bharat Shah.
\newblock Time-series forecasting of seasonal items sales using machine learning {\textendash} {{A}} comparative analysis.
\newblock {\em International Journal of Information Management Data Insights}, 2(1):100058, April 2022.
\newblock Available online at \url{https://linkinghub.elsevier.com/retrieve/pii/S2667096822000027}, Accessed: October 2023.

\bibitem{wanMultivariateTemporalConvolutional2019}
Renzhuo Wan, Shuping Mei, Jun Wang, Min Liu, and Fan Yang.
\newblock Multivariate {{Temporal Convolutional Network}}: {{A Deep Neural Networks Approach}} for {{Multivariate Time Series Forecasting}}.
\newblock {\em Electronics}, 8(8):876, August 2019.
\newblock Available online at \url{https://www.mdpi.com/2079-9292/8/8/876}, Accessed: October 2023.

\bibitem{xiaoAFSTGCNPredictionMultivariate2022}
Yuteng Xiao, Kaijian Xia, Hongsheng Yin, Yu-Dong Zhang, Zhenjiang Qian, Zhaoyang Liu, Yuehan Liang, and Xiaodan Li.
\newblock {{AFSTGCN}}: {{Prediction}} for multivariate time series using an adaptive fused spatial-temporal graph convolutional network.
\newblock {\em Digital Communications and Networks}, July 2022.
\newblock Available online at \url{https://linkinghub.elsevier.com/retrieve/pii/S2352864822001419}, Accessed: October 2023.

\bibitem{triebeNeuralProphetExplainableForecasting2021}
Oskar Triebe, Hansika Hewamalage, Polina Pilyugina, Nikolay Laptev, Christoph Bergmeir, and Ram Rajagopal.
\newblock {{NeuralProphet}}: {{Explainable Forecasting}} at {{Scale}}, November 2021.
\newblock Available online at \url{http://arxiv.org/abs/2111.15397}, Accessed: October 2023.

\bibitem{castleRobustApproachesForecasting2015}
Jennifer~L. Castle, Michael~P. Clements, and David~F. Hendry.
\newblock Robust approaches to forecasting.
\newblock {\em International Journal of Forecasting}, 31(1):99--112, January 2015.
\newblock Available online at \url{https://linkinghub.elsevier.com/retrieve/pii/S0169207014001496}, Accessed: October 2023.

\bibitem{tyralisLargescaleAssessmentProphet2018}
Hristos Tyralis and Georgia~A. Papacharalampous.
\newblock Large-scale assessment of {{Prophet}} for multi-step ahead forecasting of monthly streamflow.
\newblock {\em Advances in Geosciences}, 45:147--153, August 2018.
\newblock Available online at \url{https://adgeo.copernicus.org/articles/45/147/2018/}, Accessed: October 2023.

\bibitem{borkinAddingAdditionalFeatures2019}
Dmitrii Borkin, Martin N{\'e}meth, German Micha{\v l}{\v c}onok, and Olga Mezentseva.
\newblock Adding {{Additional Features}} to {{Improve Time Series Prediction}}.
\newblock {\em Research Papers Faculty of Materials Science and Technology Slovak University of Technology}, 27(45):72--78, September 2019.
\newblock Available online at \url{https://www.sciendo.com/article/10.2478/rput-2019-0028}, Accessed: October 2023.

\bibitem{chenMultiScaleAdaptiveGraph2023}
Ling Chen, Donghui Chen, Zongjiang Shang, Binqing Wu, Cen Zheng, Bo~Wen, and Wei Zhang.
\newblock Multi-{{Scale Adaptive Graph Neural Network}} for {{Multivariate Time Series Forecasting}}.
\newblock {\em IEEE Transactions on Knowledge and Data Engineering}, pages 1--14, 2023.
\newblock Available online at \url{https://ieeexplore.ieee.org/document/10105527/}, Accessed: October 2023.

\bibitem{wuConnectingDotsMultivariate2020}
Zonghan Wu, Shirui Pan, Guodong Long, Jing Jiang, Xiaojun Chang, and Chengqi Zhang.
\newblock Connecting the {{Dots}}: {{Multivariate Time Series Forecasting}} with {{Graph Neural Networks}}.
\newblock In {\em Proceedings of the 26th {{ACM SIGKDD International Conference}} on {{Knowledge Discovery}} \& {{Data Mining}}}, pages 753--763, {Virtual Event CA USA}, August 2020. {ACM}.
\newblock Available online at \url{https://dl.acm.org/doi/10.1145/3394486.3403118}, Accessed: October 2023.

\bibitem{hallCorrelationbasedFeatureSelection1999}
Mark~A Hall.
\newblock Correlation-based {{Feature Selection}} for {{Machine Learning}}.
\newblock {\em University of Waikato}, April 1999.

\bibitem{jimenezMultiobjectiveEvolutionaryFeature2017}
F.~Jim{\'e}nez, G.~S{\'a}nchez, J.M. Garc{\'i}a, G.~Sciavicco, and L.~Miralles.
\newblock Multi-objective evolutionary feature selection for online sales forecasting.
\newblock {\em Neurocomputing}, 234:75--92, April 2017.
\newblock Available online at \url{https://linkinghub.elsevier.com/retrieve/pii/S0925231216315612}, Accessed: October 2023.

\bibitem{yangNetworkTrafficForecasting2021}
Hanyu Yang, Xutao Li, Wenhao Qiang, Yuhan Zhao, Wei Zhang, and Chang Tang.
\newblock A network traffic forecasting method based on {{SA}} optimized {{ARIMA}}{\textendash}{{BP}} neural network.
\newblock {\em Computer Networks}, 193:108102, July 2021.
\newblock Available online at \url{https://linkinghub.elsevier.com/retrieve/pii/S1389128621001821}, Accessed: October 2023.

\bibitem{hyndman2018forecasting}
Rob~J Hyndman and George Athanasopoulos.
\newblock {\em Forecasting: Principles and Practice}.
\newblock {OTexts}, 2018.

\bibitem{boxTimeSeriesAnalysis1976}
{\relax George E.P}.~Box and Gwilym~M. Jenkins.
\newblock {\em Time {{Series Analysis}}: {{Forecasting}} and {{Control}}}.
\newblock Holden-Day, 1976.

\bibitem{xiaoDualStageAttention2021}
Yuteng Xiao, Hongsheng Yin, Yudong Zhang, Honggang Qi, Yundong Zhang, and Zhaoyang Liu.
\newblock A dual-stage attention-based {{Conv}}-{{LSTM}} network for spatio-temporal correlation and multivariate time series prediction.
\newblock {\em International Journal of Intelligent Systems}, 36(5):2036--2057, May 2021.
\newblock Available online at \url{https://onlinelibrary.wiley.com/doi/10.1002/int.22370}, Accessed: October 2023.

\bibitem{parmezanEvaluationStatisticalMachine2019}
Antonio Rafael~Sabino Parmezan, Vinicius~M.A. Souza, and Gustavo~E.A.P.A. Batista.
\newblock Evaluation of statistical and machine learning models for time series prediction: {{Identifying}} the state-of-the-art and the best conditions for the use of each model.
\newblock {\em Information Sciences}, 484:302--337, May 2019.
\newblock Available online at \url{https://linkinghub.elsevier.com/retrieve/pii/S0020025519300945}, Accessed: October 2023.

\bibitem{limTimeseriesForecastingDeep2021}
Bryan Lim and Stefan Zohren.
\newblock Time-series forecasting with deep learning: A survey.
\newblock {\em Philosophical Transactions of the Royal Society A: Mathematical, Physical and Engineering Sciences}, 379(2194):20200209, April 2021.
\newblock Available online at \url{https://royalsocietypublishing.org/doi/10.1098/rsta.2020.0209}, Accessed: October 2023.

\bibitem{wilmsInterpretableVectorAutoRegressions2017}
Ines Wilms, Sumanta Basu, Jacob Bien, and David~S. Matteson.
\newblock Interpretable {{Vector AutoRegressions}} with {{Exogenous Time Series}}, November 2017.
\newblock Available online at \url{http://arxiv.org/abs/1711.03623}, Accessed: October 2023.

\bibitem{bojerKaggleForecastingCompetitions2021}
Casper~Solheim Bojer and Jens~Peder Meldgaard.
\newblock Kaggle forecasting competitions: {{An}} overlooked learning opportunity.
\newblock {\em International Journal of Forecasting}, 37(2):587--603, April 2021.
\newblock Available online at \url{https://linkinghub.elsevier.com/retrieve/pii/S0169207020301114}, Accessed: October 2023.

\bibitem{wolpertNoFreeLunch1997}
D.H. Wolpert and W.G. Macready.
\newblock No free lunch theorems for optimization.
\newblock {\em IEEE Transactions on Evolutionary Computation}, 1(1):67--82, April 1997.
\newblock Available online at \url{http://ieeexplore.ieee.org/document/585893/}, Accessed: October 2023.

\bibitem{hongProbabilisticElectricLoad2016}
Tao Hong and Shu Fan.
\newblock Probabilistic electric load forecasting: {{A}} tutorial review.
\newblock {\em International Journal of Forecasting}, 32(3):914--938, July 2016.
\newblock Available online at \url{https://linkinghub.elsevier.com/retrieve/pii/S0169207015001508}, Accessed: October 2023.

\bibitem{SARIMAXIntroduction2023}
{Statsmodels} {Python Package} - {{SARIMAX}}, September 2023.
\newblock Available online at \url{https://www.statsmodels.org/dev/examples/notebooks/generated/statespace\_sarimax\_stata.html}, Accessed: October 2023.

\bibitem{triebeNeuralProphet2021}
{NeuralProphet} {Python Package}, January 2021.
\newblock Available online at \url{https://neuralprophet.com/contents.html}, Accessed: October 2023.

\bibitem{tranSelectionSignificantInput2015}
H.D. Tran, N.~Muttil, and B.J.C. Perera.
\newblock Selection of significant input variables for time series forecasting.
\newblock {\em Environmental Modelling \& Software}, 64:156--163, February 2015.
\newblock Available online at \url{https://linkinghub.elsevier.com/retrieve/pii/S1364815214003442}, Accessed: October 2023.

\bibitem{nokeriForecastingUsingARIMA2021}
Tshepo~Chris Nokeri.
\newblock {\em Forecasting {{Using ARIMA}}, {{SARIMA}}, and the {{Additive Model}}}, pages 21--50.
\newblock {Apress}, {Berkeley, CA}, 2021.
\newblock Available online at \url{https://link.springer.com/10.1007/978-1-4842-7110-0\_2}, Accessed: October 2023.

\bibitem{moroffMachineLearningStatistics2021}
Nikolas~Ulrich Moroff, Ersin Kurt, and Josef Kamphues.
\newblock Machine {{Learning}} and {{Statistics}}: {{A Study}} for assessing innovative {{Demand Forecasting Models}}.
\newblock {\em Procedia Computer Science}, 180:40--49, 2021.
\newblock Available online at \url{https://linkinghub.elsevier.com/retrieve/pii/S1877050921001654}, Accessed: October 2023.

\bibitem{vagropoulosComparisonSARIMAXSARIMA2016}
Stylianos~I. Vagropoulos, G.~I. Chouliaras, E.~G. Kardakos, C.~K. Simoglou, and A.~G. Bakirtzis.
\newblock Comparison of {{SARIMAX}}, {{SARIMA}}, modified {{SARIMA}} and {{ANN-based}} models for short-term {{PV}} generation forecasting.
\newblock In {\em 2016 {{IEEE International Energy Conference}} ({{ENERGYCON}})}, pages 1--6, {Leuven, Belgium}, April 2016. {IEEE}.
\newblock Available online at \url{http://ieeexplore.ieee.org/document/7514029/}, Accessed: October 2023.

\bibitem{ServicesMonthlyData}
Services - monthly data.
\newblock \url{https://ec.europa.eu/eurostat/databrowser/view/ei\_bsse\_m\_r2/default/table?lang=en}.
\newblock Database: Eurostat - Statistical Office of the European Union, Accessed: October 2023.

\bibitem{MetadataBusinessConsumer}
Metadata of {{Business}} and consumer surveys (ei\_bcs).
\newblock \url{https://ec.europa.eu/eurostat/cache/metadata/en/ei_bcs_esms.htm}.
\newblock Database: Eurostat - Statistical Office of the European Union, Accessed: October 2023.

\bibitem{taylorForecastingScale2018}
Sean~J. Taylor and Benjamin Letham.
\newblock Forecasting at {{Scale}}.
\newblock {\em The American Statistician}, 72(1):37--45, January 2018.
\newblock Available online at \url{https://www.tandfonline.com/doi/full/10.1080/00031305.2017.1380080}, Accessed: October 2023.

\bibitem{triebeARNetSimpleAutoRegressive2019}
Oskar Triebe, Nikolay Laptev, and Ram Rajagopal.
\newblock {{AR-Net}}: {{A}} simple {{Auto-Regressive Neural Network}} for time-series, November 2019.
\newblock Available online at \url{http://arxiv.org/abs/1911.12436}, Accessed: October 2023.

\bibitem{CORTESIBANEZ2020385}
Juan~Antonio {Cort{\'e}s-Ib{\'a}{\~n}ez}, Sergio Gonz{\'a}lez, Jos{\'e}~Javier {Valle-Alonso}, Juli{\'a}n Luengo, Salvador Garc{\'i}a, and Francisco Herrera.
\newblock Preprocessing methodology for time series: {{An}} industrial world application case study.
\newblock {\em Information Sciences}, 514:385--401, 2020.
\newblock Available online at \url{https://www.sciencedirect.com/science/article/pii/S002002551931076X}, Accessed: October 2023.

\bibitem{Eurostat}
Eurostat database - statistical office of the european union.
\newblock \url{https://ec.europa.eu/eurostat/de/}.

\bibitem{EurostatPythonPackage}
Eurostat {{Python Package}}.
\newblock \url{https://pypi.org/project/eurostat}.

\bibitem{kotuDataScienceProcess2019}
Vijay Kotu and Bala Deshpande.
\newblock Data {{Science Process}}.
\newblock In {\em Data {{Science}}}, pages 19--37. {Elsevier}, 2019.
\newblock Available online at \url{https://linkinghub.elsevier.com/retrieve/pii/B9780128147610000022}, Accessed: October 2023.

\bibitem{assafExplainableDeepNeural2019}
Roy Assaf and Anika Schumann.
\newblock Explainable {{Deep Neural Networks}} for {{Multivariate Time Series Predictions}}.
\newblock In {\em Proceedings of the {{Twenty-Eighth International Joint Conference}} on {{Artificial Intelligence}}}, pages 6488--6490, {Macao, China}, August 2019. {International Joint Conferences on Artificial Intelligence Organization}.
\newblock Available online at \url{https://www.ijcai.org/proceedings/2019/932}, Accessed: October 2023.

\bibitem{gogtayPrinciplesCorrelationAnalysis2017}
{\relax NJ}~Gogtay and {\relax UM}~Thatte.
\newblock Principles of {{Correlation Analysis}}.
\newblock {\em Journal of The Association of Physicians of India}, 65, 2017.
\newblock Available online at \url{https://www.kem.edu/wp-content/uploads/2012/06/9-Principles_of_correlation-1.pdf}, Accessed: October 2023.

\bibitem{muthukrishnanLASSOFeatureSelection2016}
R~Muthukrishnan and R~Rohini.
\newblock {{LASSO}}: {{A}} feature selection technique in predictive modeling for machine learning.
\newblock In {\em 2016 {{IEEE International Conference}} on {{Advances}} in {{Computer Applications}} ({{ICACA}})}, pages 18--20, {Coimbatore, India}, October 2016. {IEEE}.
\newblock Available online at \url{http://ieeexplore.ieee.org/document/7887916/}, Accessed: October 2023.

\bibitem{ververidisSequentialForwardFeature2005}
Dimitrios Ververidis and Constantine Kotropoulos.
\newblock Sequential forward feature selection with low computational cost.
\newblock In {\em 13th {{European Signal Processing Conference}}}, pages 1--4, {Antalya, Turkey}, 2005.
\newblock Available online at \url{https://ieeexplore.ieee.org/abstract/document/7078140}, Accessed: October 2023.

\bibitem{Pandas}
Pandas {{Python Package}}.
\newblock \url{https://pandas.pydata.org/docs/index.html}.

\bibitem{Allam2022}
Zaheer Allam, Simon~Elias Bibri, and Samantha~A. Sharpe.
\newblock The rising impacts of the {{COVID-19}} pandemic and the {{Russia}}\&ndash;{{Ukraine}} war: {{Energy}} transition, climate justice, global inequality, and supply chain disruption.
\newblock {\em Resources}, 11(99), 2022.
\newblock Available online at \url{https://www.mdpi.com/2079-9276/11/11/99}, Accessed: October 2023.

\bibitem{MonthlyCruideOil}
Monthly cruide oil supply.
\newblock \url{https://ec.europa.eu/eurostat/databrowser/view/nrg\_cb\_cosm/default/table?lang=en}.
\newblock Database: Eurostat - Statistical Office of the European Union, Accessed: October 2023.

\bibitem{CrudeOilImports}
Crude oil imports by field of production - monthly data.
\newblock \url{https://ec.europa.eu/eurostat/databrowser/view/nrg\_ti\_coifpm/default/table?lang=en}.
\newblock Database: Eurostat - Statistical Office of the European Union, Accessed: October 2023.

\bibitem{ImportPricesIndustry}
Import prices in industry - monthly data.
\newblock \url{https://ec.europa.eu/eurostat/databrowser/view/sts\_inpi\_m/default/table?lang=en}.
\newblock Database: Eurostat - Statistical Office of the European Union, Accessed: October 2023.

\bibitem{FoodPriceMonitoring}
Food price monitoring tool.
\newblock \url{https://ec.europa.eu/eurostat/databrowser/view/prc\_fsc\_idx/default/table?lang=en}.
\newblock Database: Eurostat - Statistical Office of the European Union, Accessed: October 2023.

\bibitem{MetadataFoodPrice}
Metadata for {{Food}} price monitoring tool (prc\_fsc\_idx).
\newblock \url{https://ec.europa.eu/eurostat/cache/metadata/en/prc\_fsc\_idx\_esms.htm}.
\newblock Database: Eurostat - Statistical Office of the European Union, Accessed: October 2023.

\bibitem{EMUConvergenceCriterion}
{{EMU}} convergence criterion series - monthly data.
\newblock \url{https://ec.europa.eu/eurostat/databrowser/view/irt\_lt\_mcby\_m/default/table?lang=en}.
\newblock Database: Eurostat - Statistical Office of the European Union, Accessed: October 2023.

\bibitem{MetadataMaastrichtCriterion}
Metadata for {{Maastricht}} criterion interest rates (irt\_lt\_mcby).
\newblock \url{https://ec.europa.eu/eurostat/cache/metadata/en/irt\_lt\_mcby\_esms.htm}.
\newblock Database: Eurostat - Statistical Office of the European Union, Accessed: October 2023.

\bibitem{RetailSaleMonthly}
Retail sale - monthly data.
\newblock \url{https://ec.europa.eu/eurostat/databrowser/view/ei\_bsrt\_m\_r2/default/table?lang=en}.
\newblock Database: Eurostat - Statistical Office of the European Union, Accessed: October 2023.

\bibitem{InterestRatesMonthly}
Interest rates - monthly data.
\newblock \url{https://ec.europa.eu/eurostat/databrowser/view/ei\_mfir\_m/default/table?lang=en}.
\newblock Database: Eurostat - Statistical Office of the European Union, Accessed: October 2023.

\bibitem{MetadataMonetaryFinancial}
Metadata of {{Monetary}} and financial indicators (ei\_mf).
\newblock \url{https://ec.europa.eu/eurostat/cache/metadata/en/ext\_go\_agg\_esms.htm}.
\newblock Database: Eurostat - Statistical Office of the European Union, Accessed: October 2023.

\bibitem{MetadataBalancePayments}
Metadata for {{Balance}} of payments (ei\_bp).
\newblock \url{https://ec.europa.eu/eurostat/cache/metadata/en/ei\_bp\_esms.htm}.
\newblock Database: Eurostat - Statistical Office of the European Union, Accessed: October 2023.

\bibitem{FinancialAccountMonthly}
Financial account - monthly data.
\newblock \url{https://ec.europa.eu/eurostat/databrowser/view/ei\_bpm6fa\_m/default/table?lang=en}.
\newblock Database: Eurostat - Statistical Office of the European Union, Accessed: October 2023.

\bibitem{ConstructionMonthlyDatab}
Construction - monthly data - index (2015 = 100) ({{NACE Rev}}. 2).
\newblock \url{https://ec.europa.eu/eurostat/databrowser/view/ei\_isbu\_m/default/table?lang=en}.
\newblock Database: Eurostat - Statistical Office of the European Union, Accessed: October 2023.

\bibitem{MetadataIndustryTrade}
Metadata for {{Industry}}, trade and services (ei\_is).
\newblock \url{https://ec.europa.eu/eurostat/cache/metadata/en/ei\_is\_esms.htm}.
\newblock Database: Eurostat - Statistical Office of the European Union, Accessed: October 2023.

\bibitem{RetailTradeMonthly}
Retail trade - monthly data - growth rates ({{NACE Rev}}. 2).
\newblock \url{https://ec.europa.eu/eurostat/databrowser/view/ei\_isrr\_m/default/table?lang=en}.
\newblock Database: Eurostat - Statistical Office of the European Union, Accessed: October 2023.

\bibitem{TurnoverVolumeSales}
Turnover and volume of sales in wholesale and retail trade - monthly data.
\newblock \url{https://ec.europa.eu/eurostat/databrowser/view/sts\_trtu\_m/default/table?lang=en}.
\newblock Database: Eurostat - Statistical Office of the European Union, Accessed: October 2023.

\bibitem{MetadataShorttermBusiness}
Metadata for {{Short-term}} business statistics.
\newblock \url{https://ec.europa.eu/eurostat/cache/metadata/en/sts\_esms.htm}.
\newblock Database: Eurostat - Statistical Office of the European Union, Accessed: October 2023.

\bibitem{ProducerPricesIndustry}
Producer prices in industry, total - monthly data.
\newblock \url{https://ec.europa.eu/eurostat/databrowser/view/sts\_inpp\_m/default/table?lang=en}.
\newblock Database: Eurostat - Statistical Office of the European Union, Accessed: October 2023.

\bibitem{EnergyMonthlyData}
Energy - monthly data.
\newblock \url{https://ec.europa.eu/eurostat/databrowser/view/ei\_isen\_m/default/table?lang=en}.
\newblock Database: Eurostat - Statistical Office of the European Union, Accessed: October 2023.

\bibitem{ArrivalsTouristAccommodation}
Arrivals at tourist accommodation establishments - monthly data.
\newblock \url{https://ec.europa.eu/eurostat/databrowser/view/tour\_occ\_arm/default/table?lang=en}.
\newblock Database: Eurostat - Statistical Office of the European Union, Accessed: October 2023.

\bibitem{MetadataOccupancyTourist}
Metadata for {{Occupancy}} of tourist accommodation establishments (tour\_occ).
\newblock \url{https://ec.europa.eu/eurostat/cache/metadata/en/tour\_occ\_esms.htm}.
\newblock Database: Eurostat - Statistical Office of the European Union, Accessed: October 2023.

\bibitem{IndustryMonthlyData}
Industry - monthly data.
\newblock \url{https://ec.europa.eu/eurostat/databrowser/view/ei\_bsin\_m\_r2/default/table?lang=en}.
\newblock Database: Eurostat - Statistical Office of the European Union, Accessed: October 2023.

\bibitem{ProducerPricesIndustrya}
Producer prices in industry, domestic market - monthly data.
\newblock \url{https://ec.europa.eu/eurostat/databrowser/view/sts\_inppd\_m/default/table?lang=en}.
\newblock Database: Eurostat - Statistical Office of the European Union, Accessed: October 2023.

\bibitem{ProducerPricesIndustryb}
Producer prices in industry, non domestic market - monthly data.
\newblock \url{https://ec.europa.eu/eurostat/databrowser/view/sts\_inppnd\_m/default/table?lang=en}.
\newblock Database: Eurostat - Statistical Office of the European Union, Accessed: October 2023.

\end{thebibliography}
\end{multicols}

\end{document}